\documentclass[superscriptaddress,twocolumn,prb,showkeys,showpacs]{revtex4-1}

\usepackage[utf8]{inputenc}
\usepackage{amsmath}   
\usepackage{graphicx}   
\usepackage{verbatim}  
\usepackage{color}     
\usepackage{subfigure}  
\usepackage{hyperref}   
\usepackage{natbib}
\usepackage{gensymb}
\usepackage{color}
\usepackage{epstopdf}
\usepackage{multirow}
\usepackage{xfrac}
\begin{document}

\title{Unveiling the complete dispersion of the giant Rashba split surface states of ferroelectric $\alpha$-GeTe(111) by alkali doping}

\author{G. Kremer}
\altaffiliation{Corresponding author.\\ geoffroy.kremer@unifr.ch}
\affiliation{D{\'e}partement de Physique and Fribourg Center for Nanomaterials, Universit{\'e} de Fribourg, CH-1700 Fribourg, Switzerland}

\author{T. Jaouen}
\affiliation{D{\'e}partement de Physique and Fribourg Center for Nanomaterials, Universit{\'e} de Fribourg, CH-1700 Fribourg, Switzerland}

\author{B. Salzmann}
\affiliation{D{\'e}partement de Physique and Fribourg Center for Nanomaterials, Universit{\'e} de Fribourg, CH-1700 Fribourg, Switzerland}

\author{L. Nicolaï}
\affiliation{New Technologies-Research Center University of West Bohemia, Plzen, Czech Republic}

\author{M. Rumo}
\affiliation{D{\'e}partement de Physique and Fribourg Center for Nanomaterials, Universit{\'e} de Fribourg, CH-1700 Fribourg, Switzerland}

\author{C. W. Nicholson}
\affiliation{D{\'e}partement de Physique and Fribourg Center for Nanomaterials, Universit{\'e} de Fribourg, CH-1700 Fribourg, Switzerland}

\author{B.~Hildebrand}
\affiliation{D{\'e}partement de Physique and Fribourg Center for Nanomaterials, Universit{\'e} de Fribourg, CH-1700 Fribourg, Switzerland}

\author{J. H. Dil}
\affiliation{Institute of physics, Ecole Polytechnique F{\'e}d{\'e}rale de Lausanne, CH-1015 Lausanne, Switzerland}
\affiliation{Photon Science Division, Paul Scherrer Institut, CH-5232 Villigen, Switzerland}

\author{J. Min{\'ar}}
\affiliation{New Technologies-Research Center University of West Bohemia, Plzen, Czech Republic}

\author{G. Springholz}
\affiliation{Institut f{\"u}r Halbleiter-und Festk{\"o}rperphysik, Johannes Kepler Universit{\"a}t, A-4040 Linz, Austria}

\author{J. Krempask{\'y}}
\affiliation{Photon Science Division, Paul Scherrer Institut, CH-5232 Villigen, Switzerland}

\author{C. Monney}
\affiliation{D{\'e}partement de Physique and Fribourg Center for Nanomaterials, Universit{\'e} de Fribourg, CH-1700 Fribourg, Switzerland}

\begin{abstract}

$\alpha$-GeTe(111) is a non-centrosymmetric ferroelectric material, for which a strong spin-orbit interaction gives rise to giant Rashba split states in the bulk and at the surface. The detailed dispersions of the surface states inside the bulk band gap remains an open question because they are located in the unoccupied part of the electronic structure, making them inaccessible to static angle-resolved photoemission spectroscopy. We show that this difficulty can be overcome via \textit{in-situ} potassium doping of the surface, leading to a rigid shift of 80 meV of the surface states into the occupied states. Thus, we resolve in great detail their dispersion and highlight their crossing at the $\bar{\Gamma}$ point, which, in comparison with density functional theory calculations, definitively confirms the Rashba mechanism. 

\end{abstract}
\date{\today}
\maketitle

%%%%%%%%%%%%%%%%%% INTRODUCTION %%%%%%%%%%%%%%%%%%%%%

\begin{center}
\textbf{I. INTRODUCTION}
\end{center}

The electronic band structure in solids with both inversion and time-reversal symmetries (TRS) is at least doubly degenerate with respect to the spin degree of freedom. In systems where the inversion symmetry is broken, this degeneracy can be lifted by the spin-orbit interaction (SOI). Rashba and Bychkov have theoretically described this phenomenon by considering a two-dimensional electron gas subject to an external out-of-plane electric field \cite{bychkov1984properties}. The resulting energy levels are split by a linear term in momentum $\mathbf{k}$ : $E^{\pm}(\mathbf{k}) =~ E_{0}~ +~ (\hbar^{2} \mathbf{k}^{2} / 2 m^{\star}) ~ \pm ~ \alpha_{R} ~ |\mathbf{k}|$,  where $m^{\star}$ is the effective mass of the electrons and  $\alpha_{R}$ is the Rashba parameter reflecting the magnitude of this so called Rashba effect. The Rashba effect was first observed by angle-resolved photoemission spectroscopy (ARPES) at the surface of Au(111), where the Shockley state shows a momentum-dependent energy splitting \cite{lashell1996spin}. Subsequently, spin-resolved ARPES confirmed its predicted in-plane spin polarisation \cite{hoesch2004spin}. Similar experimental investigations on other surfaces and interfaces, with variable splitting magnitude, have been performed \cite{roten1999,hoch2002,koro2004,krupin2005,suga2006,cercellier2006interplay,ast2007giant,hirahara2008origin,hsieh2009observation,dil2009spin,sakamoto2009peculiar,varykhalov2012ir,tamai2013spin,krasovskii2015spin,zhang2017,yaji2018}.  More recently, work has been done on BiTeX (X = I, Cl, Br) polar semiconductors, where the first evidence of giant bulk Rashba split states was observed by ARPES and spin-resolved ARPES \cite{bahramy2011origin,ishizaka2011giant,crepaldi2012giant,landolt2012disentanglement,sakano2012three,landolt2013bulk,landolt2015direct}
.

In recent decades, an increasing amount of research has been carried out aiming to enhance the control of spin polarised currents in nanostructured materials. Rashba systems have appeared as an ideal playground to address these concepts and have been recently used as an efficient spin to charge converter in prototype spintronics devices \cite{sanchez2013spin,lesne2016highly,oyarzun2016evidence}. The synthesis of new materials with giant and tunable Rashba split electronic states, adjustable position of the chemical potential and, ideally, spin-polarised states crossing the Fermi level ($E_{F}$), is highly desirable in the view of future applications in the growing area of spintronics \cite{soumyanarayanan2016emergent}.

In this context, $\alpha$-GeTe(111) is a promising compound \cite{rinaldi2016evidence}. It is a ferroelectric semiconductor which exhibits spin-polarised bulk and surface split electronic states with the largest Rashba parameter currently reported, and has been termed a ferroelectric Rashba semiconductor (FERS) \cite{di2013electric,picozzi2014ferroelectric}. Bulk and surface states are Rashba split due to  out-of-plane inversion symmetry breaking and large SOI. 

Considerable work has been carried out to experimentally characterise the electronic band structure of \hbox{$\alpha$-GeTe(111)}, in particular its bulk Rashba split states \cite{liebmann2016giant,krempasky2016entanglement,krempasky2016disentangling}. Spin-resolved ARPES measurements have experimentally confirmed the link between the spin texture and the ferroelectricity, demonstrating that the spin-polarised band structure can be reversibly manipulated either by modifying the surface termination\cite{rinaldi2018ferroelectric} or with an electric field \cite{krempasky2018operando}. Since ARPES experiments can only probe the occupied part of the band structure, these studies have mainly focused on the bulk states, and the surface states still lack a complete and direct experimental characterisation. In particular, their detailed dispersion at the $\bar{\Gamma}$ point of the Brillouin zone (BZ) remains unknown because the states are located in the unoccupied part of the band structure. Direct measurements of the surface states is an important point to address in order to fully understand the electronic band structure of the system, and to test the validity of current theories.

In the present work, we show that the surface states of $\alpha$-GeTe(111) can be shifted down to $E_{F}$ and experimentally addressed with ARPES via \textit{in-situ} surface deposition of potassium (K). Their complete dispersion is resolved by taking advantage of the electronic thermal occupation at room temperature (RT). In particular, we resolve their crossing point (CP) at $\bar{\Gamma}$, thus confirming the TRS conservation and the Rashba picture. We further demonstrate their non-parabolic dispersion, in excellent agreement with density functional theory (DFT) band structure calculations. We show that the Rashba-type splitting is unaffected by the electron doping, allowing us to extrapolate the energy position of the surface states on the bare $\alpha$-GeTe(111).

\bigskip

%%%%%%%%%%%%%%%%%% METHODS %%%%%%%%%%%%%%%%%%%%%

\begin{center}
\textbf{II. METHODS}
\end{center}

 Two hundred nanometers-thick $\alpha$-GeTe(111) films were grown by molecular beam epitaxy (MBE) on InP(111) substrates and then characterised in a different experimental setup, coupling a surface preparation chamber, with low-temperature scanning tunneling microscopy (LT-STM), X-ray photoemission spectroscopy (XPS) and ARPES techniques. To avoid surface degradation and oxidation during the transfer between the two systems, a protective stack of amorphous Te and Se capping layers was deposited \textit{in-situ} after MBE growth and subsequently removed in ultrahigh vacuum (UHV) in the STM/XPS/ARPES system in a base pressure of 7 $\times$ 10$^{-11}$ mbar at 500 K. Note that in contrast to (Bi$_{1-x}$Sb$_{x}$)$_{2}$Te$_{3}$ \cite{kremer2019recovery}, for $\alpha$-GeTe(111) a pure Te cap does not provide satisfactory results. 

The low-temperature (4.5 K) STM images were obtained using an Omicron LT-STM, in a pressure better than 5 $\times$ 10$^{-11}$ mbar. The low-energy electron diffraction (LEED)  patterns were recorded with a SPECS ErLEED at RT and 80 eV. ARPES measurements were performed using a Scienta DA30 photoelectron analyser with monochromatised HeI$_{\alpha}$ radiation ($h \nu$ = 21.2 eV, SPECS UVLS with TMM 304 monochromator) and at RT, if not further specified. Energy and angular resolutions were better than 10 meV and 0.1°,  respectively. XPS measurements were carried out with a monochromatised Al K$\alpha$ source (SPECS $\mu-$FOCUS, resolution better than 300 meV).  The K evaporations were performed at RT from a SAES getter source in a pressure better than 5 $\times$ 10$^{-10}$ mbar.

 \textit{Ab-initio} calculations are based on DFT as implemented in the fully relativistic spin-polarized relativistic Korringa-Kohn-Rostoker (SPR-KKR) theory \cite{ebert2011calculating}. The electronic structure of a semi-infinite surface of $\alpha$-GeTe is described, including all relativistic  effects, by the Dirac equation, which is solved using screened KKR formalism. The local density approximation based potentials were treated within the atomic sphere approximation and for the multipole expansion of the Green function an angular momentum cutoff $l_{max}$=3 was used.  The structural of Te-terminated surface has been taken from the structural relaxation presented in the work of \hbox{Krempask{\'y} \textit{et al.} \cite{krempasky2018operando}}. The electronic structure of semi-infinite $\alpha$-GeTe is represented by means of Bloch spectral function (BSF), \textit{i.e.} imaginary part of the KKR Green function.

\bigskip

%%%%%%%%%%%%%%%%%%%%% DICUSSSIONS %%%%%%%%%%%%%%%%%%%

\begin{figure}[t]
\includegraphics[scale=0.28]{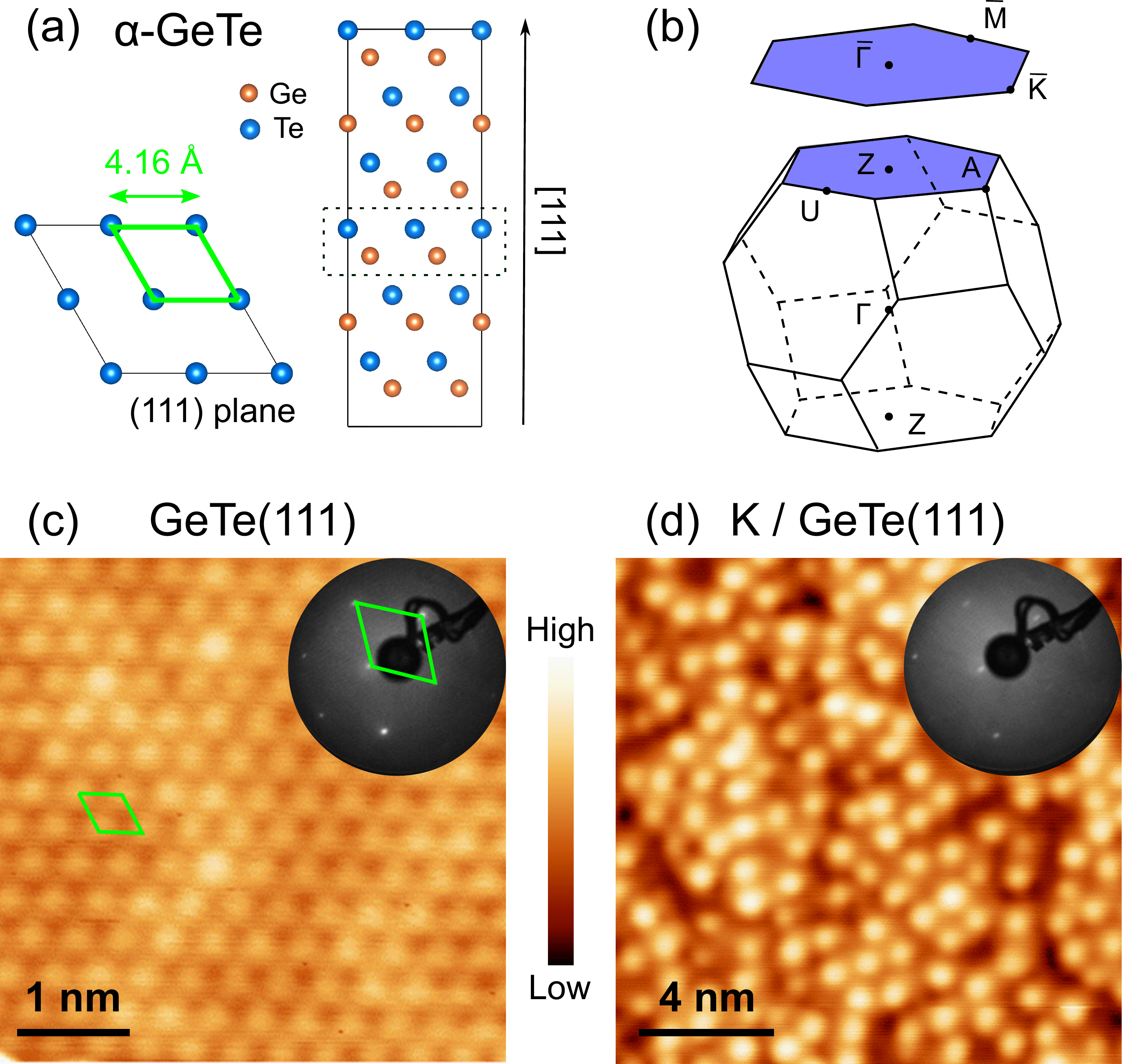}
\caption{(a) Crystal structure of $\alpha$-GeTe along the [111] crystallographic direction (right) and the corresponding projection in the (111) atomic plane (bottom left). The two-dimensional (2D) unit cell (green rhombus) parameter is extracted from X-ray diffraction measurements \cite{bauer2013lattice}. (b) Corresponding three-dimensional (3D) Brillouin zone (BZ) and its 2D projection in the (111) plane (blue). (c) STM image of the bare $\alpha$-GeTe(111) surface (U = -1.5 V, I = 0.15 nA) and associated LEED pattern in the top right corner. The 2D unit cells in real and reciprocal space are  indicated in green.  (d) STM image of the $\alpha$-GeTe(111) surface covered by 0.09~ML of K (U = -1.5 V, I = 0.15 nA) and the associated LEED pattern in the top right corner.}
\label{fig1}
\end{figure}

\bigskip

\begin{center}
\textbf{III. RESULTS}
\end{center} 

The crystal structure of $\alpha$-GeTe(111) is shown in the right part of Fig. \ref{fig1}(a) (space group \textit{R3m}). It corresponds to a stacking of "bilayers" \textit{i.e.} sequences of Te and Ge planes (see black-dashed rectangle) along the rhombohedral [111] crystallographic direction. The distance between the planes of Te and Ge within the same "bilayer" is smaller than the one between two "bilayers". This large rhomboedral lattice distorsion, which is equal to about 10 $\%$ of the lattice parameter\cite{krempasky2016disentangling}, provides the ferroelectric order. The (111) cut of $\alpha$-GeTe corresponds either to a Te or Ge terminated plane, leading either to an upward or a downward ferroelectric polarisation at the surface. In the following, we address the band dispersion of the surface states of a Te-terminated \hbox{$\alpha$-GeTe(111)} surface, which is predicted to be energetically most favorable \cite{deringer2012ab}.

\begin{figure*}[t]
\includegraphics[width=160mm]{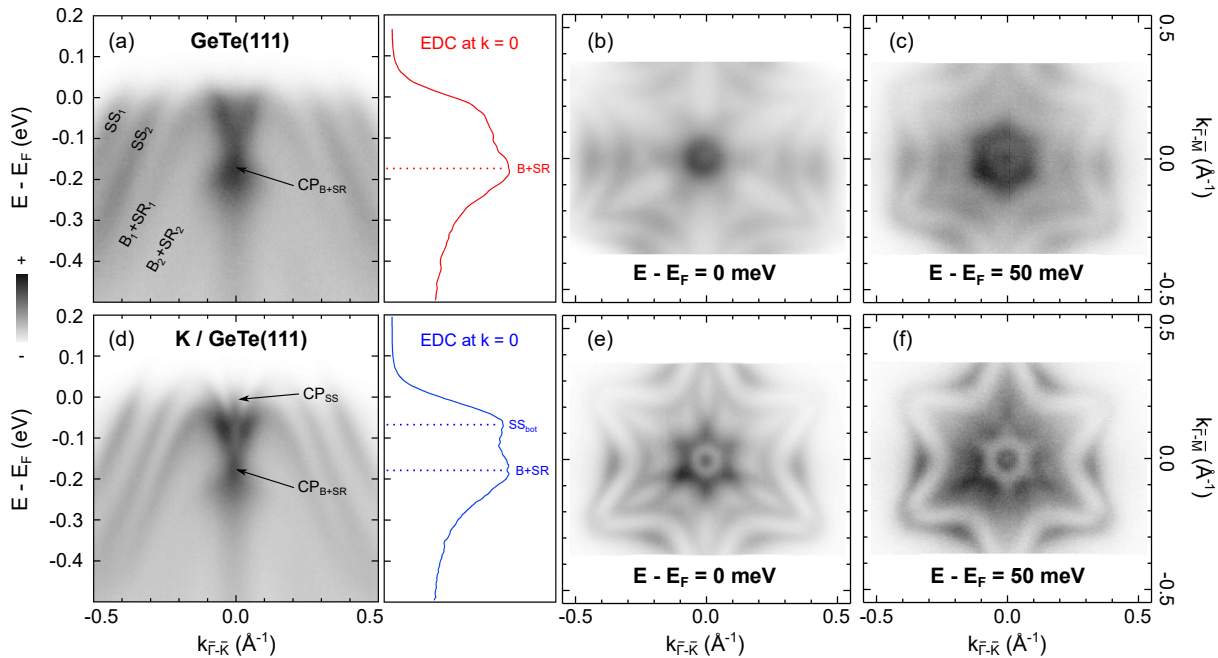}
\caption{ARPES spectra along the $\bar{K}-\bar{\Gamma}-\bar{K}$ high-symmetry line of the 2D BZ (left) and the corresponding energy-distribution curve (EDC) taken at $k = 0$ (right) for (a) bare  $\alpha$-GeTe(111) and (d) K/$\alpha$-GeTe(111). Associated constant energy surfaces taken at $E - E_{F} = 0$ meV and $E - E_{F} = 50$ meV for (b,c) bare  $\alpha$-GeTe(111) and (e,f) K/$\alpha$-GeTe(111).}
\label{fig2}
\end{figure*}

To this end, we have annealed our capped sample at 500 K in order to recover a fresh surface. The resulting surface has first been characterised by STM and LEED, as shown in Fig. \ref{fig1}(c). This panel shows an atomically-resolved STM image of the (111) surface of $\alpha$-GeTe. It reveals hexagonally well-arranged atoms with an interatomic distance estimated as (4.1~$\pm$~0.2) \AA , in good agreement with the STM image simulated by DFT calculations of a pristine Te-terminated surface of $\alpha$-GeTe(111) in the work of Deringer \textit{et al.} \cite{deringer2012ab}. Furthermore, the LEED measurements exhibit a sharp ($1 \times 1$) hexagonal pattern, demonstrating the long-range surface order and the expected crystal symmetry after the desorption procedure. The chemical composition of our Te-terminated $\alpha$-GeTe(111) surface  was also cross-checked by XPS measurements that show negligible surface contaminants, sharp Ge $2p$ and Te $3d$ core levels and only residual traces of Se capping layer remaining from the decapping procedure (see Fig. \ref{fig4} in the Appendix).

The electronic band structure of $\alpha$-GeTe(111) measured by ARPES is shown in Fig. \ref{fig2}(a-c). ARPES intensity along the $\bar{\Gamma}-\bar{K}$ direction (see the 2D BZ in Fig. \ref{fig1}(b)) exhibits all the spectroscopic signatures of a clean Te-terminated surface. Due to spectral broadening, the bulk states B$_{1}$ and B$_{2}$ are smeared out with their surface resonances SR$_{1}$ and SR$_{2}$ \cite{krempasky2016disentangling,PhysResearchKrempasky2020}. We also resolve the surface states SS$_{1}$ and SS$_{2}$, forming characteristic "snowflake" constant energy surfaces at \hbox{$E - E_{F} = 0$ meV} (Fig. \ref{fig2}(b)) and $E - E_{F} = 50$ meV (Fig. \ref{fig2}(c)), in very good agreement with measurements and calculations for Te-terminated $\alpha$-GeTe(111) \cite{krempasky2016disentangling,rinaldi2018ferroelectric,PhysResearchKrempasky2020}. Based on our STM, LEED, XPS and ARPES results, we conclude that our $\alpha$-GeTe(111) surface is very well ordered and Te-terminated,  both at the atomic and macroscopic scales.

Subsequently, we have deposited K atoms on the  \hbox{$\alpha$-GeTe(111)} surface. Fig. \ref{fig1}(d) displays the corresponding STM image where the K adatoms are clearly distinguishable as large and bright protrusions. The coverage is estimated as $(0.09 \pm 0.01)$ monolayer (ML) (one ML corresponds to one K atom per alpha-GeTe(111) surface unit cell). A ($1 \times 1$) hexagonal pattern is still visible in LEED but with a more pronounced diffuse background. This background is due to the randomly organised K atoms at the surface, without any coherent surface reconstruction, as visible in the STM image. The ($1 \times 1$) pattern in LEED is due to the $\alpha$-GeTe(111) sublayer, which is not distinguishable in the STM topography.

\begin{figure}[t]
\includegraphics[scale=0.069]{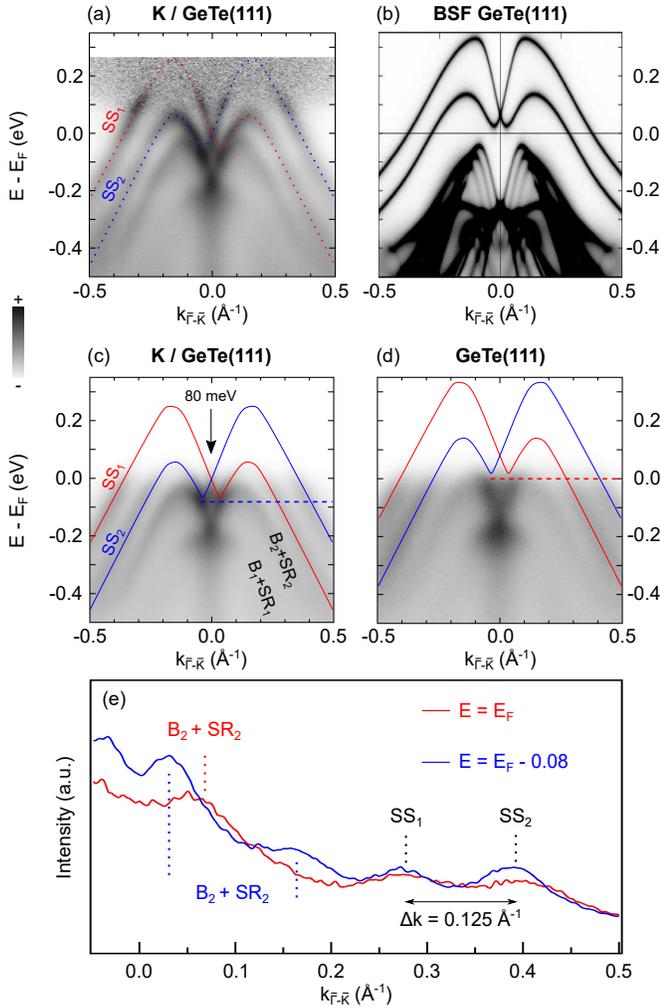}
\caption{(a)  Fermi-Dirac distribution divided (at RT) ARPES spectrum along the $\bar{K}-\bar{\Gamma}-\bar{K}$ high-symmetry line of the 2D BZ for K/$\alpha$-GeTe(111) surface. Dashed-blue and red lines correspond to guides to the eye showing the full dispersions of SS$_{1}$ and SS$_{2}$. (b) BSF calculations of Te-terminated \hbox{$\alpha$-GeTe(111)}. (c,d) ARPES spectra of K/$\alpha$-GeTe(111) and \hbox{$\alpha$-GeTe(111)}. Solid red and blue lines are shifted by 80 meV up to $E_{F}$ in (d) compared to panel (a). (e) Momentum distribution curves (MDCs) associated to the horizontal colored-dashed lines in panels (c,d).}
\label{fig3}
\end{figure}

Let us now discuss the evolution of the electronic band structure after K adsorption. A freshly K-doped surface with identical doping and spectroscopic features to the one corresponding to 0.09 ML of K was obtained after an annealing at 500 K of a K-saturated $\alpha$-GeTe(111) surface. This point is discussed in the Fig. \ref{fig5} of the Appendix, in particular by means of work function measurements. The electronic band structure of K/$\alpha$-GeTe(111) along the $\bar{\Gamma}-\bar{K}$ high-symmetry direction (see Fig. \ref{fig2}(d)) exhibits sharp dispersions where bulk states are unchanged but the surface states SS$_{1}$ and SS$_{2}$ are rigidly shifted down to $E_{F}$ with respect to the undoped surface. The upper part of the SR$_{1}$+B$_{1}$ and SR$_{2}$+B$_{2}$ branches is also slightly shifted to high binding energy, as visible around $k = -  \thinspace 0.1 \thinspace $\AA$^{-1}$. This is likely due to a small shift of the surface resonances which are now quasi-degenerate with the bulk states close to $E_{F}$, in contrary to the \hbox{$\alpha$-GeTe(111)} case \cite{PhysResearchKrempasky2020}. In the case of bare $\alpha$-GeTe(111), SS$_{1}$ and SS$_{2}$ respectively cross $E_{F}$ at $k_{F,1} = -  \thinspace 0.395 \thinspace $\AA$^{-1}$ and $k_{F,2} = -  \thinspace 0.270 \thinspace $\AA$^{-1}$ whereas for K/$\alpha$-GeTe(111), $k_{F,1} = -  \thinspace 0.350 \thinspace $\AA$^{-1}$ and $k_{F,2} = -  \thinspace 0.225 \thinspace $\AA$^{-1}$. This demonstrates  that for such a low coverage the \textit{Rashba-type splitting is unaffected}, since the momentum splitting $\Delta k = k_{F,1}-k_{F,2}$ is not changed upon K doping. The energy shift of the Rashba split surface states towards higher binding energy is also visible from the Fermi surface (see Fig. \ref{fig2}(e)) where the two outer arcs are localised at lower momenta compared to the undoped surface, both along the $\bar{\Gamma}-\bar{K}$ and $\bar{\Gamma}-\bar{M}$ high-symmetry directions. Through K deposition, we therefore electron-dope the $\alpha$-GeTe(111) surface, leading to a rigid energy shift of the surface states down to the occupied states\cite{hossain2008situ,boyle2019topological}.

A direct consequence of the rigid energy shift of the surface electronic structure is the emergence of CP$_{SS}$ at normal emission and $E_{F}$, related to pure surface states, as highlighted by a black arrow in Fig. \ref{fig2}(d). It corresponds to the crossing of  SS$_{1}$ and SS$_{2}$ which are degenerate at $k = 0$, as expected in a Rashba scenario due to the TRS conservation (Kramers degeneracy theorem). This CP$_{SS}$ is not visible for the bare $\alpha$-GeTe(111), because it occurs at too high energy in the unoccupied part of the band structure. We have plotted energy distribution curves (EDCs) taken at the $\bar{\Gamma}$ point, \textit{i.e.} at $k = 0$, of the bare and K-doped $\alpha$-GeTe(111) surfaces in the right-hand side of Fig. \ref{fig2}(a) and Fig. \ref{fig2}(d), respectively. In the case of $\alpha$-GeTe(111), the EDC exhibits one dominant contribution, associated to the CP$_{B+SR}$ of bulk and surface resonance states, also denoted by a black arrow in Fig. \ref{fig2}(a). In the K-doped case, there are two contributions: again the CP$_{B+SR}$ but also the bottom of the surface states. The CP$_{SS}$  is not distinguishable in the EDC because of the Fermi edge cut-off. As we will see below, it is possible to overcome this by dividing by the Fermi-Dirac distribution. Finally, the Fermi surface map (see Fig. \ref{fig2}(e)) is accordingly modified: some spectral weight appears exactly at the center of the "snowflake"  at  $\bar{\Gamma}$, forming a single point. Looking above $E_{F}$,  \textit{e.g.} at $E_{F} + 50$ meV (see Fig. \ref{fig2}(f)), this  point becomes a ring corresponding to the lifting of degeneracy of the surface states away from the $\bar{\Gamma}$ point. Thus the experimental band dispersions summarized in Fig. \ref{fig2} indicate that, in contrast to earlier studies \cite{liebmann2016giant}, surface states are not degenerate with bulk states at the $\bar{\Gamma}$ point. Moreover, the CP$_{B+SR}$ are not affected by surface electronic doping because they remain located around 180 meV below $E_{F}$ as previously reported \cite{krempasky2016disentangling}.

Taking advantage of the well-defined band dispersions obtained at RT after K adsorption, we next divide the raw data of Fig. \ref{fig2}(d) by the Fermi-Dirac distribution, a well known procedure used to gain access to the thermally occupied band structure in the range of a few tens of meV above $E_{F}$. It is particularly pertinent at RT because the width of the Fermi edge is approximately equal to 100 meV.  Fig. \ref{fig3}(a) displays the corresponding result. The detailed dispersion of the surface states is now evident, in particular their CP at $E_{F}$ and their linear dispersion in the vicinity of the $\bar{\Gamma}$ point. Guides to the eye are represented as red and blue dashed lines to highlight the two spin-polarised surface states SS$_{1}$ and SS$_{2}$. In the present case, low-temperature measurements are not beneficial because they reduce our access to the thermally occupied part of the band structure (see \hbox{Fig. \ref{fig6} in the Appendix}).

In order to obtain the energy position of the CP$_{SS}$ in bare $\alpha$-GeTe(111), Fig. \ref{fig3}(c,d) show ARPES spectra for K/GeTe(111) and the bare surface with guides to the eye respectively shifted by 0 and +80 meV in comparison to panel (a). For this size of  rigid shift, the guides to the eye also fits very well to the surface states in the bare surface case. To confirm this, we have extracted momentum distribution curves (MDCs) for pertinent binding energies in Fig. \ref{fig3}(e).  The MDCs are plotted at $E_{F}$ for $\alpha$-GeTe(111) (red line) and at 80 meV below $E_{F}$ in the K/$\alpha$-GeTe(111) case (blue line). They exhibit similar surface states contributions, with a comparable Rashba-type splitting ($\Delta k = \thinspace 0.125 \thinspace $\AA$^{-1}$), but different bulk/surface resonance ones. Indeed, whereas the MDC associated to the bare surface (red line) shows only one broad contribution centered at $k = +  \thinspace 0.062 \thinspace $\AA$^{-1}$ for B$_{2}$+SR$_{2}$, the one of the K-doped surface (blue line) exhibits two components, respectively centered at \hbox{$k = + \thinspace 0.030 \thinspace $\AA$^{-1}$} and $k = +  \thinspace 0.165 \thinspace $\AA$^{-1}$. This not only demonstrates that upon K adsorption the bulk band structure is not modified but also shows that the effect of K is to rigidly shift the surface states down to the $E_{F}$ by 80 meV, value obtained by matching the momentum positions of SS$_{1}$ and SS$_{2}$ in the blue and red MDCs. Overall, our analysis allows us to obtain the energy position of the surface states in bare $\alpha$-GeTe(111) and conclude that their CP is positionned around 80 meV above $E_{F}$. 

Finally, our observations are in excellent agreement with BSF calculations performed on a Te-terminated \hbox{$\alpha$-GeTe(111)} surface, as shown in Fig. \ref{fig3}(b). Indeed, the dispersion of the surface states is well reproduced by the simulations, especially their "V shape" above the CP at  $\bar{\Gamma}$  and their non-parabolic character, remarkably different from nearly free electrons dispersion. The energy position of the \hbox{CP$_{SS}$} is theoretically estimated to be 70 meV above $E_{F}$, in very good agreement with our experimental findings.

\bigskip

\begin{center}
\textbf{IV. CONCLUSIONS}
\end{center}

In summary, direct measurement of the Rashba split surface states of $\alpha$-GeTe(111) has been experimentally realised thanks to K doping, in excellent agreement with state-of-the-art band structure calculations. We report the detailed dispersion of these states in the vicinity of the $\bar{\Gamma}$ point, confirming the existence of their CP in the bulk band gap and by consequence the associated Rashba scenario with TRS conservation. We also present the definitive proof that the CP of the surface states is located in the unoccupied part of the band structure, a finding which were not properly etablished from previous experimental and theoretical works \cite{liebmann2016giant,rinaldi2018ferroelectric}. Furthermore, we show that the surface states shift in energy upon K doping, but \textit{not} the bulk states. By consequence, we can conclude that the shift of the surface states is not the result of band bending and that they are decoupled from the bulk states. We also find that K doping does not affect their Rashba splitting. These results clearly prove that the \textit{intrinsic} origin of the giant Rashba splitting of the surface states of $\alpha$-GeTe(111) is largely arising from the inversion symmetry breaking in the bulk. If it would be merely a surface effect, the Rashba splitting of the surface states should significantly change upon surface doping. Our work is therefore an important step for the understanding of the electronic band structure of $\alpha$-GeTe(111), which is a promising material for multi-functional spintronics devices. It opens the way for future investigations of the full spin texture of the surface states via spin-resolved ARPES around $\bar{\Gamma}$. Finally, our results call for additional studies of both the unoccupied band structure and the electron dynamics of the system \textit{via} time-resolved ARPES measurements, in particular the study of the potential relaxation channels of the electrons from the conduction band to the surface states.

\bigskip

%%%%%%%%%%%%%%%%%% REMERCIEMENTS %%%%%%%%%%%%%%%%%%%%%

\begin{center}
\textbf{ACKNOWLEDGMENTS}
\end{center}

This project was supported by the Swiss National Science Foundation (SNSF) Grant \hbox{No. P00P2$\_$170597} and the Austrian Science Funds (FWF), Project No. P30860-N27. L.N. and J.M. would like to thank CEDAMNF project financed by the Ministry of Education, Youth and Sports of Czech Repuplic, Project No. CZ.02.1.01/0.0/0.0/15$\_$003/0000358 and Czech Science Foundation (GACR), Proj. 20-18725S. We are very grateful to P. Aebi for fruitful discussions and for sharing with us his photoemission setup. Skillful technical assistance was provided by F. Bourqui, \hbox{B. Hediger} and O. Raetzo.

%%%%%%%%%%%%%%%%%% APPENDIX %%%%%%%%%%%%%%%%%%%%%

\bigskip

\begin{center}
\textbf{APPENDIX}
\end{center}

\begin{center}
\textbf{1. X-ray photoemission spectroscopy}
\end{center}

Fig. \ref{fig4} displays the XPS spectrum of the bare \hbox{$\alpha$-GeTe(111)} surface, obtained after the desorption of the capping layer. It mainly exhibits the Ge $2p$ and Te $3d$ core levels, with weak secondary electrons and contaminants signals, demonstrating the quality of the surface.

\begin{figure}[h!]
\centering
\includegraphics[width=1 \columnwidth]{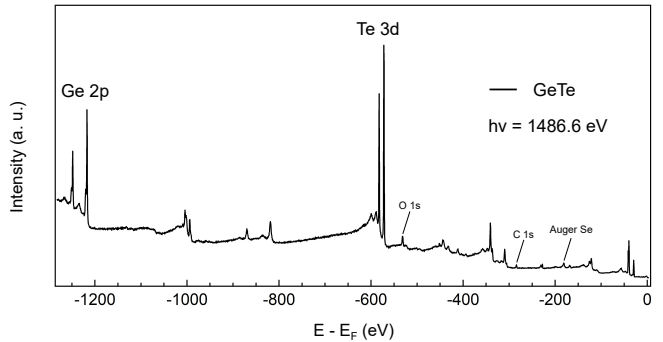}
\caption{XPS spectrum of the $\alpha$-GeTe(111) surface obtained after the initial desorption procedure to remove the capping layer. The photon energy is $h \nu$ = 1486.6 eV.}
\label{fig4}
\end{figure}

\begin{center}
\textbf{2. Extended data and sample characterisation}
\end{center}

Hereafter, we discuss how we have obtained the sample corresponding to the ARPES data on the K/$\alpha$-GeTe(111) surface presented in the main text. We started from a bare $\alpha$-GeTe(111) sample which has been characterised with LEED, ARPES, work function measurements and XPS. The corresponding data are shown in Fig. \ref{fig5}. LEED, ARPES and XPS have been discussed in the main text. The work function measurement of the material is done by applying a bias of -8 V and measuring the energy cut-off of the secondary electrons ($E_{vac}$). The work function of the material is given by the formula: $\Phi = h\nu - ( E_{F} - E_{vac} )$, with $E_{F}$ defined as the kinetic energy of the electrons at the Fermi level. By doing such a measurement on $\alpha$-GeTe(111), we extract a value of 4.7 eV, well above the K bulk value (1.8 eV). So, we are expecting an electron transfer from the K adatoms to the $\alpha$-GeTe(111) surface.

\begin{figure}[h!]
\centering
\includegraphics[width=1\columnwidth]{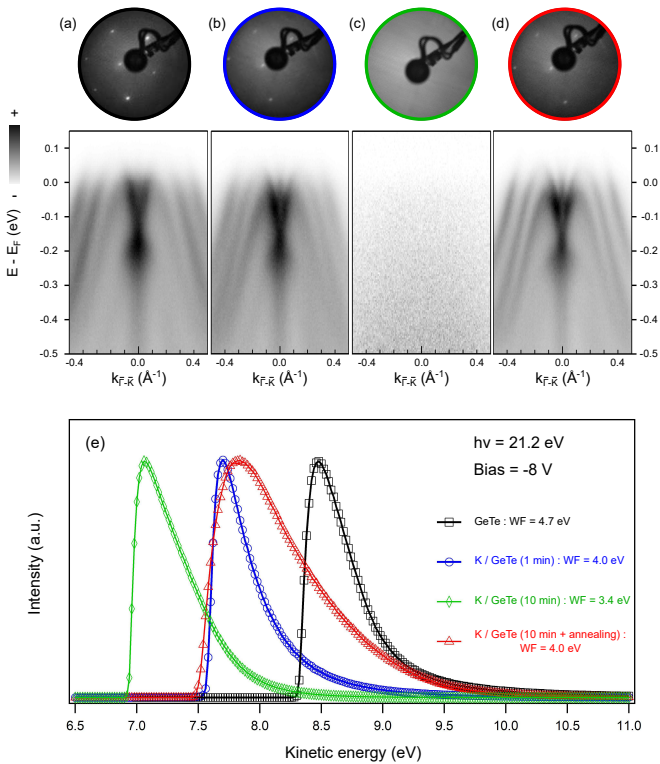}
\caption{Series of LEED patterns (top) and ARPES spectra (bottom) along the $\bar{K}-\bar{\Gamma}-\bar{K}$ high-symmetry line of the 2D BZ for (a) $\alpha$-GeTe(111), (b) $\alpha$-GeTe(111) with 1 min of K deposition, (c) $\alpha$-GeTe(111) with 10 min of K deposition and (d) $\alpha$-GeTe(111) with 10 min of deposition followed by an annealing at 500 K. Note that the spectra from panels (a) and (d) have been measured with better statistics and with a smaller energy step compared to panels (b) and (c). (e) Corresponding low-energy cut-off of the secondary electrons as probed by HeI$_{\alpha}$ photoemission by applying a voltage bias of -8 V to the sample surface.}
\label{fig5}
\end{figure}

After 1 min of potassium deposition (0.09 ML estimated using STM), the LEED pattern is more diffuse and the ARPES data exhibit a shift of 80 meV of the surface states down to the Fermi level, without any modification of the bulk states (Fig. \ref{fig5}(b)). As expected, we observe a reduction of the work function down to \hbox{4.0 eV} (Fig. \ref{fig5}(e)).  Increasing the amount of potassium at the surface by a factor of ten, \textit{i.e.} 10 min of deposition, leads to the loss of both the LEED spots and the ARPES band structure (Fig. \ref{fig5}(c)), and to a decrease of the work function down to 3.4 eV.

By further annealing at 500 K, we recover the same LEED and ARPES data as those obtained after 1 min  of deposition (Fig. \ref{fig5}(d)) and work function value of 4.0 eV. This demonstrates that the annealing procedure allows for recovering an electron doping level of the \hbox{K-saturated} $\alpha$-GeTe(111) surface identical to the one obtained after 1 min of K deposition. 

\begin{center}
\textbf{3. ARPES as a function of the temperature}
\end{center}

To show the temperature dependence of the band structure of $\alpha$-GeTe(111) and K/$\alpha$-GeTe(111), Fig. \ref{fig6} displays ARPES measurements taken at RT and at \hbox{T = 80 K.} As it can be seen, the dispersions of the surface, bulk and surface resonance states are the same. The only effects of lowering the temperature are to reduce both the spectral broadening and the width of the energy region in the thermally occupied part of the band structure that can be probed thanks to the thermal broadening of the Fermi edge.

\begin{figure}[h!]
\centering
\includegraphics[width=1\columnwidth]{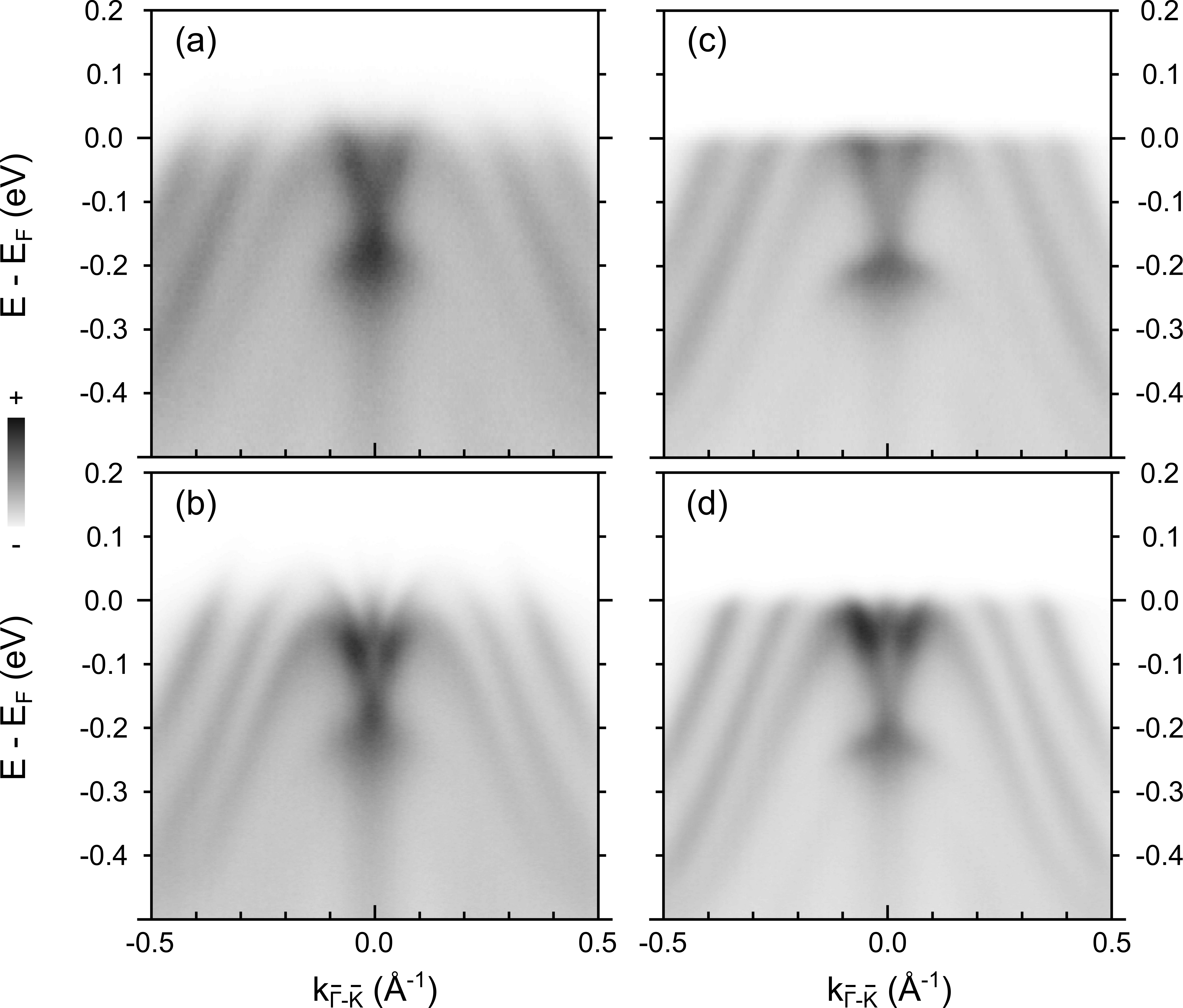}
\caption{Temperature-dependent ARPES measurements along the $\bar{K}-\bar{\Gamma}-\bar{K}$ high-symmetry line of the 2D BZ, at RT for (a) $\alpha$-GeTe(111) and (b) K/$\alpha$-GeTe(111) and at T = 80 K for (c)   $\alpha$-GeTe(111) and (d) K/$\alpha$-GeTe(111).}
\label{fig6}
\end{figure}


\begin{thebibliography}{45}%
\makeatletter
\providecommand \@ifxundefined [1]{%
 \@ifx{#1\undefined}
}%
\providecommand \@ifnum [1]{%
 \ifnum #1\expandafter \@firstoftwo
 \else \expandafter \@secondoftwo
 \fi
}%
\providecommand \@ifx [1]{%
 \ifx #1\expandafter \@firstoftwo
 \else \expandafter \@secondoftwo
 \fi
}%
\providecommand \natexlab [1]{#1}%
\providecommand \enquote  [1]{``#1''}%
\providecommand \bibnamefont  [1]{#1}%
\providecommand \bibfnamefont [1]{#1}%
\providecommand \citenamefont [1]{#1}%
\providecommand \href@noop [0]{\@secondoftwo}%
\providecommand \href [0]{\begingroup \@sanitize@url \@href}%
\providecommand \@href[1]{\@@startlink{#1}\@@href}%
\providecommand \@@href[1]{\endgroup#1\@@endlink}%
\providecommand \@sanitize@url [0]{\catcode `\\12\catcode `\$12\catcode
  `\&12\catcode `\#12\catcode `\^12\catcode `\_12\catcode `\%12\relax}%
\providecommand \@@startlink[1]{}%
\providecommand \@@endlink[0]{}%
\providecommand \url  [0]{\begingroup\@sanitize@url \@url }%
\providecommand \@url [1]{\endgroup\@href {#1}{\urlprefix }}%
\providecommand \urlprefix  [0]{URL }%
\providecommand \Eprint [0]{\href }%
\providecommand \doibase [0]{http://dx.doi.org/}%
\providecommand \selectlanguage [0]{\@gobble}%
\providecommand \bibinfo  [0]{\@secondoftwo}%
\providecommand \bibfield  [0]{\@secondoftwo}%
\providecommand \translation [1]{[#1]}%
\providecommand \BibitemOpen [0]{}%
\providecommand \bibitemStop [0]{}%
\providecommand \bibitemNoStop [0]{.\EOS\space}%
\providecommand \EOS [0]{\spacefactor3000\relax}%
\providecommand \BibitemShut  [1]{\csname bibitem#1\endcsname}%
\let\auto@bib@innerbib\@empty
%</preamble>
\bibitem [{\citenamefont {Bychkov}\ and\ \citenamefont
  {Rashba}(1984)}]{bychkov1984properties}%
  \BibitemOpen
  \bibfield  {author} {\bibinfo {author} {\bibfnamefont {Y.~A.}\ \bibnamefont
  {Bychkov}}\ and\ \bibinfo {author} {\bibfnamefont {E.~I.}\ \bibnamefont
  {Rashba}},\ }\href@noop {} {\bibfield  {journal} {\bibinfo  {journal} {JETP
  lett.}\ }\textbf {\bibinfo {volume} {39}},\ \bibinfo {pages} {78} (\bibinfo
  {year} {1984})}\BibitemShut {NoStop}%
\bibitem [{\citenamefont {LaShell}\ \emph {et~al.}(1996)\citenamefont
  {LaShell}, \citenamefont {McDougall},\ and\ \citenamefont
  {Jensen}}]{lashell1996spin}%
  \BibitemOpen
  \bibfield  {author} {\bibinfo {author} {\bibfnamefont {S.}~\bibnamefont
  {LaShell}}, \bibinfo {author} {\bibfnamefont {B.~A.}\ \bibnamefont
  {McDougall}}, \ and\ \bibinfo {author} {\bibfnamefont {E.}~\bibnamefont
  {Jensen}},\ }\href@noop {} {\bibfield  {journal} {\bibinfo  {journal} {Phys.
  Rev. Lett.}\ }\textbf {\bibinfo {volume} {77}},\ \bibinfo {pages} {3419}
  (\bibinfo {year} {1996})}\BibitemShut {NoStop}%
\bibitem [{\citenamefont {Hoesch}\ \emph {et~al.}(2004)\citenamefont {Hoesch},
  \citenamefont {Muntwiler}, \citenamefont {Petrov}, \citenamefont
  {Hengsberger}, \citenamefont {Patthey}, \citenamefont {Shi}, \citenamefont
  {Falub}, \citenamefont {Greber},\ and\ \citenamefont
  {Osterwalder}}]{hoesch2004spin}%
  \BibitemOpen
  \bibfield  {author} {\bibinfo {author} {\bibfnamefont {M.}~\bibnamefont
  {Hoesch}}, \bibinfo {author} {\bibfnamefont {M.}~\bibnamefont {Muntwiler}},
  \bibinfo {author} {\bibfnamefont {V.~N.}\ \bibnamefont {Petrov}}, \bibinfo
  {author} {\bibfnamefont {M.}~\bibnamefont {Hengsberger}}, \bibinfo {author}
  {\bibfnamefont {L.}~\bibnamefont {Patthey}}, \bibinfo {author} {\bibfnamefont
  {M.}~\bibnamefont {Shi}}, \bibinfo {author} {\bibfnamefont {M.}~\bibnamefont
  {Falub}}, \bibinfo {author} {\bibfnamefont {T.}~\bibnamefont {Greber}}, \
  and\ \bibinfo {author} {\bibfnamefont {J.}~\bibnamefont {Osterwalder}},\
  }\href@noop {} {\bibfield  {journal} {\bibinfo  {journal} {Phys. Rev. B}\
  }\textbf {\bibinfo {volume} {69}},\ \bibinfo {pages} {241401(R)} (\bibinfo
  {year} {2004})}\BibitemShut {NoStop}%
\bibitem [{\citenamefont {Rotenberg}\ \emph {et~al.}(1999)\citenamefont
  {Rotenberg}, \citenamefont {Chung},\ and\ \citenamefont {Kevan}}]{roten1999}%
  \BibitemOpen
  \bibfield  {author} {\bibinfo {author} {\bibfnamefont {E.}~\bibnamefont
  {Rotenberg}}, \bibinfo {author} {\bibfnamefont {J.~W.}\ \bibnamefont
  {Chung}}, \ and\ \bibinfo {author} {\bibfnamefont {S.~D.}\ \bibnamefont
  {Kevan}},\ }\href@noop {} {\bibfield  {journal} {\bibinfo  {journal} {Phys.
  Rev. Lett.}\ }\textbf {\bibinfo {volume} {82}},\ \bibinfo {pages} {4066}
  (\bibinfo {year} {1999})}\BibitemShut {NoStop}%
\bibitem [{\citenamefont {Hochstrasser}\ \emph {et~al.}(2002)\citenamefont
  {Hochstrasser}, \citenamefont {Tobin}, \citenamefont {Rotenberg},\ and\
  \citenamefont {Kevan}}]{hoch2002}%
  \BibitemOpen
  \bibfield  {author} {\bibinfo {author} {\bibfnamefont {M.}~\bibnamefont
  {Hochstrasser}}, \bibinfo {author} {\bibfnamefont {J.~G.}\ \bibnamefont
  {Tobin}}, \bibinfo {author} {\bibfnamefont {E.}~\bibnamefont {Rotenberg}}, \
  and\ \bibinfo {author} {\bibfnamefont {S.~D.}\ \bibnamefont {Kevan}},\
  }\href@noop {} {\bibfield  {journal} {\bibinfo  {journal} {Phys. Rev. Lett.}\
  }\textbf {\bibinfo {volume} {89}},\ \bibinfo {pages} {216802} (\bibinfo
  {year} {2002})}\BibitemShut {NoStop}%
\bibitem [{\citenamefont {Koroteev}\ \emph {et~al.}(2004)\citenamefont
  {Koroteev}, \citenamefont {Bihlmayer}, \citenamefont {Gayone}, \citenamefont
  {Chulkov}, \citenamefont {Bl{\"u}gel}, \citenamefont {Echenique},\ and\
  \citenamefont {Hofmann}}]{koro2004}%
  \BibitemOpen
  \bibfield  {author} {\bibinfo {author} {\bibfnamefont {Y.~M.}\ \bibnamefont
  {Koroteev}}, \bibinfo {author} {\bibfnamefont {G.}~\bibnamefont {Bihlmayer}},
  \bibinfo {author} {\bibfnamefont {J.~E.}\ \bibnamefont {Gayone}}, \bibinfo
  {author} {\bibfnamefont {E.~V.}\ \bibnamefont {Chulkov}}, \bibinfo {author}
  {\bibfnamefont {S.}~\bibnamefont {Bl{\"u}gel}}, \bibinfo {author}
  {\bibfnamefont {P.~M.}\ \bibnamefont {Echenique}}, \ and\ \bibinfo {author}
  {\bibfnamefont {P.}~\bibnamefont {Hofmann}},\ }\href@noop {} {\bibfield
  {journal} {\bibinfo  {journal} {Phys. Rev. Lett.}\ }\textbf {\bibinfo
  {volume} {93}},\ \bibinfo {pages} {046403} (\bibinfo {year}
  {2004})}\BibitemShut {NoStop}%
\bibitem [{\citenamefont {Krupin}\ \emph {et~al.}(2005)\citenamefont {Krupin},
  \citenamefont {Bihlmayer}, \citenamefont {Starke}, \citenamefont {Gorovikov},
  \citenamefont {Prieto}, \citenamefont {D{\"o}brich}, \citenamefont
  {Bl{\"u}gel},\ and\ \citenamefont {Kaindl}}]{krupin2005}%
  \BibitemOpen
  \bibfield  {author} {\bibinfo {author} {\bibfnamefont {O.}~\bibnamefont
  {Krupin}}, \bibinfo {author} {\bibfnamefont {G.}~\bibnamefont {Bihlmayer}},
  \bibinfo {author} {\bibfnamefont {K.}~\bibnamefont {Starke}}, \bibinfo
  {author} {\bibfnamefont {S.}~\bibnamefont {Gorovikov}}, \bibinfo {author}
  {\bibfnamefont {J.~E.}\ \bibnamefont {Prieto}}, \bibinfo {author}
  {\bibfnamefont {K.}~\bibnamefont {D{\"o}brich}}, \bibinfo {author}
  {\bibfnamefont {S.}~\bibnamefont {Bl{\"u}gel}}, \ and\ \bibinfo {author}
  {\bibfnamefont {G.}~\bibnamefont {Kaindl}},\ }\href@noop {} {\bibfield
  {journal} {\bibinfo  {journal} {Phys. Rev. B}\ }\textbf {\bibinfo {volume}
  {71}},\ \bibinfo {pages} {201403(R)} (\bibinfo {year} {2005})}\BibitemShut
  {NoStop}%
\bibitem [{\citenamefont {Sugawara}\ \emph {et~al.}(2006)\citenamefont
  {Sugawara}, \citenamefont {Sato}, \citenamefont {Souma}, \citenamefont
  {Takahashi}, \citenamefont {Arai},\ and\ \citenamefont {Sasaki}}]{suga2006}%
  \BibitemOpen
  \bibfield  {author} {\bibinfo {author} {\bibfnamefont {K.}~\bibnamefont
  {Sugawara}}, \bibinfo {author} {\bibfnamefont {T.}~\bibnamefont {Sato}},
  \bibinfo {author} {\bibfnamefont {S.}~\bibnamefont {Souma}}, \bibinfo
  {author} {\bibfnamefont {T.}~\bibnamefont {Takahashi}}, \bibinfo {author}
  {\bibfnamefont {M.}~\bibnamefont {Arai}}, \ and\ \bibinfo {author}
  {\bibfnamefont {T.}~\bibnamefont {Sasaki}},\ }\href@noop {} {\bibfield
  {journal} {\bibinfo  {journal} {Phys. Rev. Lett.}\ }\textbf {\bibinfo
  {volume} {96}},\ \bibinfo {pages} {046411} (\bibinfo {year}
  {2006})}\BibitemShut {NoStop}%
\bibitem [{\citenamefont {Cercellier}\ \emph {et~al.}(2006)\citenamefont
  {Cercellier}, \citenamefont {Didiot}, \citenamefont {Fagot-Revurat},
  \citenamefont {Kierren}, \citenamefont {Moreau}, \citenamefont {Malterre},\
  and\ \citenamefont {Reinert}}]{cercellier2006interplay}%
  \BibitemOpen
  \bibfield  {author} {\bibinfo {author} {\bibfnamefont {H.}~\bibnamefont
  {Cercellier}}, \bibinfo {author} {\bibfnamefont {C.}~\bibnamefont {Didiot}},
  \bibinfo {author} {\bibfnamefont {Y.}~\bibnamefont {Fagot-Revurat}}, \bibinfo
  {author} {\bibfnamefont {B.}~\bibnamefont {Kierren}}, \bibinfo {author}
  {\bibfnamefont {L.}~\bibnamefont {Moreau}}, \bibinfo {author} {\bibfnamefont
  {D.}~\bibnamefont {Malterre}}, \ and\ \bibinfo {author} {\bibfnamefont
  {F.}~\bibnamefont {Reinert}},\ }\href@noop {} {\bibfield  {journal} {\bibinfo
   {journal} {Phys. Rev. B}\ }\textbf {\bibinfo {volume} {73}},\ \bibinfo
  {pages} {195413} (\bibinfo {year} {2006})}\BibitemShut {NoStop}%
\bibitem [{\citenamefont {Ast}\ \emph {et~al.}(2007)\citenamefont {Ast},
  \citenamefont {Henk}, \citenamefont {Ernst}, \citenamefont {Moreschini},
  \citenamefont {Falub}, \citenamefont {Pacil{\'e}}, \citenamefont {Bruno},
  \citenamefont {Kern},\ and\ \citenamefont {Grioni}}]{ast2007giant}%
  \BibitemOpen
  \bibfield  {author} {\bibinfo {author} {\bibfnamefont {C.~R.}\ \bibnamefont
  {Ast}}, \bibinfo {author} {\bibfnamefont {J.}~\bibnamefont {Henk}}, \bibinfo
  {author} {\bibfnamefont {A.}~\bibnamefont {Ernst}}, \bibinfo {author}
  {\bibfnamefont {L.}~\bibnamefont {Moreschini}}, \bibinfo {author}
  {\bibfnamefont {M.~C.}\ \bibnamefont {Falub}}, \bibinfo {author}
  {\bibfnamefont {D.}~\bibnamefont {Pacil{\'e}}}, \bibinfo {author}
  {\bibfnamefont {P.}~\bibnamefont {Bruno}}, \bibinfo {author} {\bibfnamefont
  {K.}~\bibnamefont {Kern}}, \ and\ \bibinfo {author} {\bibfnamefont
  {M.}~\bibnamefont {Grioni}},\ }\href@noop {} {\bibfield  {journal} {\bibinfo
  {journal} {Phys. Rev. Lett.}\ }\textbf {\bibinfo {volume} {98}},\ \bibinfo
  {pages} {186807} (\bibinfo {year} {2007})}\BibitemShut {NoStop}%
\bibitem [{\citenamefont {Hirahara}\ \emph {et~al.}(2008)\citenamefont
  {Hirahara}, \citenamefont {Miyamoto}, \citenamefont {Kimura}, \citenamefont
  {Niinuma}, \citenamefont {Bihlmayer}, \citenamefont {Chulkov}, \citenamefont
  {Nagao}, \citenamefont {Matsuda}, \citenamefont {Qiao}, \citenamefont
  {Shimada} \emph {et~al.}}]{hirahara2008origin}%
  \BibitemOpen
  \bibfield  {author} {\bibinfo {author} {\bibfnamefont {T.}~\bibnamefont
  {Hirahara}}, \bibinfo {author} {\bibfnamefont {K.}~\bibnamefont {Miyamoto}},
  \bibinfo {author} {\bibfnamefont {A.}~\bibnamefont {Kimura}}, \bibinfo
  {author} {\bibfnamefont {Y.}~\bibnamefont {Niinuma}}, \bibinfo {author}
  {\bibfnamefont {G.}~\bibnamefont {Bihlmayer}}, \bibinfo {author}
  {\bibfnamefont {E.~V.}\ \bibnamefont {Chulkov}}, \bibinfo {author}
  {\bibfnamefont {T.}~\bibnamefont {Nagao}}, \bibinfo {author} {\bibfnamefont
  {I.}~\bibnamefont {Matsuda}}, \bibinfo {author} {\bibfnamefont
  {S.}~\bibnamefont {Qiao}}, \bibinfo {author} {\bibfnamefont {K.}~\bibnamefont
  {Shimada}},  \emph {et~al.},\ }\href@noop {} {\bibfield  {journal} {\bibinfo
  {journal} {New J. Phys.}\ }\textbf {\bibinfo {volume} {10}},\ \bibinfo
  {pages} {083038} (\bibinfo {year} {2008})}\BibitemShut {NoStop}%
\bibitem [{\citenamefont {Hsieh}\ \emph {et~al.}(2009)\citenamefont {Hsieh},
  \citenamefont {Xia}, \citenamefont {Wray}, \citenamefont {Qian},
  \citenamefont {Pal}, \citenamefont {Dil}, \citenamefont {Osterwalder},
  \citenamefont {Meier}, \citenamefont {Bihlmayer}, \citenamefont {Kane} \emph
  {et~al.}}]{hsieh2009observation}%
  \BibitemOpen
  \bibfield  {author} {\bibinfo {author} {\bibfnamefont {D.}~\bibnamefont
  {Hsieh}}, \bibinfo {author} {\bibfnamefont {Y.}~\bibnamefont {Xia}}, \bibinfo
  {author} {\bibfnamefont {L.}~\bibnamefont {Wray}}, \bibinfo {author}
  {\bibfnamefont {D.}~\bibnamefont {Qian}}, \bibinfo {author} {\bibfnamefont
  {A.}~\bibnamefont {Pal}}, \bibinfo {author} {\bibfnamefont {J.~H.}\
  \bibnamefont {Dil}}, \bibinfo {author} {\bibfnamefont {J.}~\bibnamefont
  {Osterwalder}}, \bibinfo {author} {\bibfnamefont {F.}~\bibnamefont {Meier}},
  \bibinfo {author} {\bibfnamefont {G.}~\bibnamefont {Bihlmayer}}, \bibinfo
  {author} {\bibfnamefont {C.~L.}\ \bibnamefont {Kane}},  \emph {et~al.},\
  }\href@noop {} {\bibfield  {journal} {\bibinfo  {journal} {Science}\ }\textbf
  {\bibinfo {volume} {323}},\ \bibinfo {pages} {919} (\bibinfo {year}
  {2009})}\BibitemShut {NoStop}%
\bibitem [{\citenamefont {Dil}(2009)}]{dil2009spin}%
  \BibitemOpen
  \bibfield  {author} {\bibinfo {author} {\bibfnamefont {J.~H.}\ \bibnamefont
  {Dil}},\ }\href@noop {} {\bibfield  {journal} {\bibinfo  {journal} {J. Phys.
  Condens. Matter}\ }\textbf {\bibinfo {volume} {21}},\ \bibinfo {pages}
  {403001} (\bibinfo {year} {2009})}\BibitemShut {NoStop}%
\bibitem [{\citenamefont {Sakamoto}\ \emph {et~al.}(2009)\citenamefont
  {Sakamoto}, \citenamefont {Kakuta}, \citenamefont {Sugawara}, \citenamefont
  {Miyamoto}, \citenamefont {Kimura}, \citenamefont {Kuzumaki}, \citenamefont
  {Ueno}, \citenamefont {Annese}, \citenamefont {Fujii}, \citenamefont {Kodama}
  \emph {et~al.}}]{sakamoto2009peculiar}%
  \BibitemOpen
  \bibfield  {author} {\bibinfo {author} {\bibfnamefont {K.}~\bibnamefont
  {Sakamoto}}, \bibinfo {author} {\bibfnamefont {H.}~\bibnamefont {Kakuta}},
  \bibinfo {author} {\bibfnamefont {K.}~\bibnamefont {Sugawara}}, \bibinfo
  {author} {\bibfnamefont {K.}~\bibnamefont {Miyamoto}}, \bibinfo {author}
  {\bibfnamefont {A.}~\bibnamefont {Kimura}}, \bibinfo {author} {\bibfnamefont
  {T.}~\bibnamefont {Kuzumaki}}, \bibinfo {author} {\bibfnamefont
  {N.}~\bibnamefont {Ueno}}, \bibinfo {author} {\bibfnamefont {E.}~\bibnamefont
  {Annese}}, \bibinfo {author} {\bibfnamefont {J.}~\bibnamefont {Fujii}},
  \bibinfo {author} {\bibfnamefont {A.}~\bibnamefont {Kodama}},  \emph
  {et~al.},\ }\href@noop {} {\bibfield  {journal} {\bibinfo  {journal} {Phys.
  Rev. Lett.}\ }\textbf {\bibinfo {volume} {103}},\ \bibinfo {pages} {156801}
  (\bibinfo {year} {2009})}\BibitemShut {NoStop}%
\bibitem [{\citenamefont {Varykhalov}\ \emph {et~al.}(2012)\citenamefont
  {Varykhalov}, \citenamefont {Marchenko}, \citenamefont {Scholz},
  \citenamefont {Rienks}, \citenamefont {Kim}, \citenamefont {Bihlmayer},
  \citenamefont {S{\'a}nchez-Barriga},\ and\ \citenamefont
  {Rader}}]{varykhalov2012ir}%
  \BibitemOpen
  \bibfield  {author} {\bibinfo {author} {\bibfnamefont {A.}~\bibnamefont
  {Varykhalov}}, \bibinfo {author} {\bibfnamefont {D.}~\bibnamefont
  {Marchenko}}, \bibinfo {author} {\bibfnamefont {M.~R.}\ \bibnamefont
  {Scholz}}, \bibinfo {author} {\bibfnamefont {E.~D.~L.}\ \bibnamefont
  {Rienks}}, \bibinfo {author} {\bibfnamefont {T.~K.}\ \bibnamefont {Kim}},
  \bibinfo {author} {\bibfnamefont {G.}~\bibnamefont {Bihlmayer}}, \bibinfo
  {author} {\bibfnamefont {J.}~\bibnamefont {S{\'a}nchez-Barriga}}, \ and\
  \bibinfo {author} {\bibfnamefont {O.}~\bibnamefont {Rader}},\ }\href@noop {}
  {\bibfield  {journal} {\bibinfo  {journal} {Phys. Rev. Lett.}\ }\textbf
  {\bibinfo {volume} {108}},\ \bibinfo {pages} {066804} (\bibinfo {year}
  {2012})}\BibitemShut {NoStop}%
\bibitem [{\citenamefont {Tamai}\ \emph {et~al.}(2013)\citenamefont {Tamai},
  \citenamefont {Meevasana}, \citenamefont {King}, \citenamefont {Nicholson},
  \citenamefont {de~la Torre}, \citenamefont {Rozbicki},\ and\ \citenamefont
  {Baumberger}}]{tamai2013spin}%
  \BibitemOpen
  \bibfield  {author} {\bibinfo {author} {\bibfnamefont {A.}~\bibnamefont
  {Tamai}}, \bibinfo {author} {\bibfnamefont {W.}~\bibnamefont {Meevasana}},
  \bibinfo {author} {\bibfnamefont {P.~D.~C.}\ \bibnamefont {King}}, \bibinfo
  {author} {\bibfnamefont {C.~W.}\ \bibnamefont {Nicholson}}, \bibinfo {author}
  {\bibfnamefont {A.}~\bibnamefont {de~la Torre}}, \bibinfo {author}
  {\bibfnamefont {E.}~\bibnamefont {Rozbicki}}, \ and\ \bibinfo {author}
  {\bibfnamefont {F.}~\bibnamefont {Baumberger}},\ }\href@noop {} {\bibfield
  {journal} {\bibinfo  {journal} {Phys. Rev. B}\ }\textbf {\bibinfo {volume}
  {87}},\ \bibinfo {pages} {075113} (\bibinfo {year} {2013})}\BibitemShut
  {NoStop}%
\bibitem [{\citenamefont {Krasovskii}(2015)}]{krasovskii2015spin}%
  \BibitemOpen
  \bibfield  {author} {\bibinfo {author} {\bibfnamefont {E.~E.}\ \bibnamefont
  {Krasovskii}},\ }\href@noop {} {\bibfield  {journal} {\bibinfo  {journal} {J.
  Phys. Condens. Matter}\ }\textbf {\bibinfo {volume} {27}},\ \bibinfo {pages}
  {493001} (\bibinfo {year} {2015})}\BibitemShut {NoStop}%
\bibitem [{\citenamefont {Zhang}\ \emph {et~al.}(2017)\citenamefont {Zhang},
  \citenamefont {Ma}, \citenamefont {Ishida}, \citenamefont {Zhao},
  \citenamefont {Xu}, \citenamefont {Lv}, \citenamefont {Yaji}, \citenamefont
  {Chen}, \citenamefont {Weng}, \citenamefont {Dai} \emph
  {et~al.}}]{zhang2017}%
  \BibitemOpen
  \bibfield  {author} {\bibinfo {author} {\bibfnamefont {P.}~\bibnamefont
  {Zhang}}, \bibinfo {author} {\bibfnamefont {J.-Z.}\ \bibnamefont {Ma}},
  \bibinfo {author} {\bibfnamefont {Y.}~\bibnamefont {Ishida}}, \bibinfo
  {author} {\bibfnamefont {L.-X.}\ \bibnamefont {Zhao}}, \bibinfo {author}
  {\bibfnamefont {Q.-N.}\ \bibnamefont {Xu}}, \bibinfo {author} {\bibfnamefont
  {B.-Q.}\ \bibnamefont {Lv}}, \bibinfo {author} {\bibfnamefont
  {K.}~\bibnamefont {Yaji}}, \bibinfo {author} {\bibfnamefont {G.-F.}\
  \bibnamefont {Chen}}, \bibinfo {author} {\bibfnamefont {H.-M.}\ \bibnamefont
  {Weng}}, \bibinfo {author} {\bibfnamefont {X.}~\bibnamefont {Dai}},  \emph
  {et~al.},\ }\href@noop {} {\bibfield  {journal} {\bibinfo  {journal} {Phys.
  Rev. Lett.}\ }\textbf {\bibinfo {volume} {118}},\ \bibinfo {pages} {046802}
  (\bibinfo {year} {2017})}\BibitemShut {NoStop}%
\bibitem [{\citenamefont {Yaji}\ \emph {et~al.}(2018)\citenamefont {Yaji},
  \citenamefont {Harasawa}, \citenamefont {Kuroda}, \citenamefont {Li},
  \citenamefont {Yan}, \citenamefont {Komori},\ and\ \citenamefont
  {Shin}}]{yaji2018}%
  \BibitemOpen
  \bibfield  {author} {\bibinfo {author} {\bibfnamefont {K.}~\bibnamefont
  {Yaji}}, \bibinfo {author} {\bibfnamefont {A.}~\bibnamefont {Harasawa}},
  \bibinfo {author} {\bibfnamefont {K.}~\bibnamefont {Kuroda}}, \bibinfo
  {author} {\bibfnamefont {R.}~\bibnamefont {Li}}, \bibinfo {author}
  {\bibfnamefont {B.}~\bibnamefont {Yan}}, \bibinfo {author} {\bibfnamefont
  {F.}~\bibnamefont {Komori}}, \ and\ \bibinfo {author} {\bibfnamefont
  {S.}~\bibnamefont {Shin}},\ }\href@noop {} {\bibfield  {journal} {\bibinfo
  {journal} {Phys. Rev. B}\ }\textbf {\bibinfo {volume} {98}},\ \bibinfo
  {pages} {041404(R)} (\bibinfo {year} {2018})}\BibitemShut {NoStop}%
\bibitem [{\citenamefont {Bahramy}\ \emph {et~al.}(2011)\citenamefont
  {Bahramy}, \citenamefont {Arita},\ and\ \citenamefont
  {Nagaosa}}]{bahramy2011origin}%
  \BibitemOpen
  \bibfield  {author} {\bibinfo {author} {\bibfnamefont {M.~S.}\ \bibnamefont
  {Bahramy}}, \bibinfo {author} {\bibfnamefont {R.}~\bibnamefont {Arita}}, \
  and\ \bibinfo {author} {\bibfnamefont {N.}~\bibnamefont {Nagaosa}},\
  }\href@noop {} {\bibfield  {journal} {\bibinfo  {journal} {Phys. Rev. B}\
  }\textbf {\bibinfo {volume} {84}},\ \bibinfo {pages} {041202(R)} (\bibinfo
  {year} {2011})}\BibitemShut {NoStop}%
\bibitem [{\citenamefont {Ishizaka}\ \emph {et~al.}(2011)\citenamefont
  {Ishizaka}, \citenamefont {Bahramy}, \citenamefont {Murakawa}, \citenamefont
  {Sakano}, \citenamefont {Shimojima}, \citenamefont {Sonobe}, \citenamefont
  {Koizumi}, \citenamefont {Shin}, \citenamefont {Miyahara}, \citenamefont
  {Kimura} \emph {et~al.}}]{ishizaka2011giant}%
  \BibitemOpen
  \bibfield  {author} {\bibinfo {author} {\bibfnamefont {K.}~\bibnamefont
  {Ishizaka}}, \bibinfo {author} {\bibfnamefont {M.~S.}\ \bibnamefont
  {Bahramy}}, \bibinfo {author} {\bibfnamefont {H.}~\bibnamefont {Murakawa}},
  \bibinfo {author} {\bibfnamefont {M.}~\bibnamefont {Sakano}}, \bibinfo
  {author} {\bibfnamefont {T.}~\bibnamefont {Shimojima}}, \bibinfo {author}
  {\bibfnamefont {T.}~\bibnamefont {Sonobe}}, \bibinfo {author} {\bibfnamefont
  {K.}~\bibnamefont {Koizumi}}, \bibinfo {author} {\bibfnamefont
  {S.}~\bibnamefont {Shin}}, \bibinfo {author} {\bibfnamefont {H.}~\bibnamefont
  {Miyahara}}, \bibinfo {author} {\bibfnamefont {A.}~\bibnamefont {Kimura}},
  \emph {et~al.},\ }\href@noop {} {\bibfield  {journal} {\bibinfo  {journal}
  {Nat. Mater.}\ }\textbf {\bibinfo {volume} {10}},\ \bibinfo {pages} {521}
  (\bibinfo {year} {2011})}\BibitemShut {NoStop}%
\bibitem [{\citenamefont {Crepaldi}\ \emph {et~al.}(2012)\citenamefont
  {Crepaldi}, \citenamefont {Moreschini}, \citenamefont {Autes}, \citenamefont
  {Tournier-Colletta}, \citenamefont {Moser}, \citenamefont {Virk},
  \citenamefont {Berger}, \citenamefont {Bugnon}, \citenamefont {Chang},
  \citenamefont {Kern} \emph {et~al.}}]{crepaldi2012giant}%
  \BibitemOpen
  \bibfield  {author} {\bibinfo {author} {\bibfnamefont {A.}~\bibnamefont
  {Crepaldi}}, \bibinfo {author} {\bibfnamefont {L.}~\bibnamefont
  {Moreschini}}, \bibinfo {author} {\bibfnamefont {G.}~\bibnamefont {Autes}},
  \bibinfo {author} {\bibfnamefont {C.}~\bibnamefont {Tournier-Colletta}},
  \bibinfo {author} {\bibfnamefont {S.}~\bibnamefont {Moser}}, \bibinfo
  {author} {\bibfnamefont {N.}~\bibnamefont {Virk}}, \bibinfo {author}
  {\bibfnamefont {H.}~\bibnamefont {Berger}}, \bibinfo {author} {\bibfnamefont
  {P.}~\bibnamefont {Bugnon}}, \bibinfo {author} {\bibfnamefont {Y.~J.}\
  \bibnamefont {Chang}}, \bibinfo {author} {\bibfnamefont {K.}~\bibnamefont
  {Kern}},  \emph {et~al.},\ }\href@noop {} {\bibfield  {journal} {\bibinfo
  {journal} {Phys. Rev. Lett.}\ }\textbf {\bibinfo {volume} {109}},\ \bibinfo
  {pages} {096803} (\bibinfo {year} {2012})}\BibitemShut {NoStop}%
\bibitem [{\citenamefont {Landolt}\ \emph {et~al.}(2012)\citenamefont
  {Landolt}, \citenamefont {Eremeev}, \citenamefont {Koroteev}, \citenamefont
  {Slomski}, \citenamefont {Muff}, \citenamefont {Neupert}, \citenamefont
  {Kobayashi}, \citenamefont {Strocov}, \citenamefont {Schmitt}, \citenamefont
  {Aliev} \emph {et~al.}}]{landolt2012disentanglement}%
  \BibitemOpen
  \bibfield  {author} {\bibinfo {author} {\bibfnamefont {G.}~\bibnamefont
  {Landolt}}, \bibinfo {author} {\bibfnamefont {S.~V.}\ \bibnamefont
  {Eremeev}}, \bibinfo {author} {\bibfnamefont {Y.~M.}\ \bibnamefont
  {Koroteev}}, \bibinfo {author} {\bibfnamefont {B.}~\bibnamefont {Slomski}},
  \bibinfo {author} {\bibfnamefont {S.}~\bibnamefont {Muff}}, \bibinfo {author}
  {\bibfnamefont {T.}~\bibnamefont {Neupert}}, \bibinfo {author} {\bibfnamefont
  {M.}~\bibnamefont {Kobayashi}}, \bibinfo {author} {\bibfnamefont {V.~N.}\
  \bibnamefont {Strocov}}, \bibinfo {author} {\bibfnamefont {T.}~\bibnamefont
  {Schmitt}}, \bibinfo {author} {\bibfnamefont {Z.~S.}\ \bibnamefont {Aliev}},
  \emph {et~al.},\ }\href@noop {} {\bibfield  {journal} {\bibinfo  {journal}
  {Phys. Rev. Lett.}\ }\textbf {\bibinfo {volume} {109}},\ \bibinfo {pages}
  {116403} (\bibinfo {year} {2012})}\BibitemShut {NoStop}%
\bibitem [{\citenamefont {Sakano}\ \emph {et~al.}(2012)\citenamefont {Sakano},
  \citenamefont {Miyawaki}, \citenamefont {Chainani}, \citenamefont {Takata},
  \citenamefont {Sonobe}, \citenamefont {Shimojima}, \citenamefont {Oura},
  \citenamefont {Shin}, \citenamefont {Bahramy}, \citenamefont {Arita} \emph
  {et~al.}}]{sakano2012three}%
  \BibitemOpen
  \bibfield  {author} {\bibinfo {author} {\bibfnamefont {M.}~\bibnamefont
  {Sakano}}, \bibinfo {author} {\bibfnamefont {J.}~\bibnamefont {Miyawaki}},
  \bibinfo {author} {\bibfnamefont {A.}~\bibnamefont {Chainani}}, \bibinfo
  {author} {\bibfnamefont {Y.}~\bibnamefont {Takata}}, \bibinfo {author}
  {\bibfnamefont {T.}~\bibnamefont {Sonobe}}, \bibinfo {author} {\bibfnamefont
  {T.}~\bibnamefont {Shimojima}}, \bibinfo {author} {\bibfnamefont
  {M.}~\bibnamefont {Oura}}, \bibinfo {author} {\bibfnamefont {S.}~\bibnamefont
  {Shin}}, \bibinfo {author} {\bibfnamefont {M.~S.}\ \bibnamefont {Bahramy}},
  \bibinfo {author} {\bibfnamefont {R.}~\bibnamefont {Arita}},  \emph
  {et~al.},\ }\href@noop {} {\bibfield  {journal} {\bibinfo  {journal} {Phys.
  Rev. B}\ }\textbf {\bibinfo {volume} {86}},\ \bibinfo {pages} {085204}
  (\bibinfo {year} {2012})}\BibitemShut {NoStop}%
\bibitem [{\citenamefont {Landolt}\ \emph {et~al.}(2013)\citenamefont
  {Landolt}, \citenamefont {Eremeev}, \citenamefont {Tereshchenko},
  \citenamefont {Muff}, \citenamefont {Slomski}, \citenamefont {Kokh},
  \citenamefont {Kobayashi}, \citenamefont {Schmitt}, \citenamefont {Strocov},
  \citenamefont {Osterwalder} \emph {et~al.}}]{landolt2013bulk}%
  \BibitemOpen
  \bibfield  {author} {\bibinfo {author} {\bibfnamefont {G.}~\bibnamefont
  {Landolt}}, \bibinfo {author} {\bibfnamefont {S.~V.}\ \bibnamefont
  {Eremeev}}, \bibinfo {author} {\bibfnamefont {O.~E.}\ \bibnamefont
  {Tereshchenko}}, \bibinfo {author} {\bibfnamefont {S.}~\bibnamefont {Muff}},
  \bibinfo {author} {\bibfnamefont {B.}~\bibnamefont {Slomski}}, \bibinfo
  {author} {\bibfnamefont {K.~A.}\ \bibnamefont {Kokh}}, \bibinfo {author}
  {\bibfnamefont {M.}~\bibnamefont {Kobayashi}}, \bibinfo {author}
  {\bibfnamefont {T.}~\bibnamefont {Schmitt}}, \bibinfo {author} {\bibfnamefont
  {V.~N.}\ \bibnamefont {Strocov}}, \bibinfo {author} {\bibfnamefont
  {J.}~\bibnamefont {Osterwalder}},  \emph {et~al.},\ }\href@noop {} {\bibfield
   {journal} {\bibinfo  {journal} {New J. Phys.}\ }\textbf {\bibinfo {volume}
  {15}},\ \bibinfo {pages} {085022} (\bibinfo {year} {2013})}\BibitemShut
  {NoStop}%
\bibitem [{\citenamefont {Landolt}\ \emph {et~al.}(2015)\citenamefont
  {Landolt}, \citenamefont {Eremeev}, \citenamefont {Tereshchenko},
  \citenamefont {Muff}, \citenamefont {Kokh}, \citenamefont {Osterwalder},
  \citenamefont {Chulkov},\ and\ \citenamefont {Dil}}]{landolt2015direct}%
  \BibitemOpen
  \bibfield  {author} {\bibinfo {author} {\bibfnamefont {G.}~\bibnamefont
  {Landolt}}, \bibinfo {author} {\bibfnamefont {S.~V.}\ \bibnamefont
  {Eremeev}}, \bibinfo {author} {\bibfnamefont {O.~E.}\ \bibnamefont
  {Tereshchenko}}, \bibinfo {author} {\bibfnamefont {S.}~\bibnamefont {Muff}},
  \bibinfo {author} {\bibfnamefont {K.~A.}\ \bibnamefont {Kokh}}, \bibinfo
  {author} {\bibfnamefont {J.}~\bibnamefont {Osterwalder}}, \bibinfo {author}
  {\bibfnamefont {E.~V.}\ \bibnamefont {Chulkov}}, \ and\ \bibinfo {author}
  {\bibfnamefont {J.~H.}\ \bibnamefont {Dil}},\ }\href@noop {} {\bibfield
  {journal} {\bibinfo  {journal} {Phys. Rev. B}\ }\textbf {\bibinfo {volume}
  {91}},\ \bibinfo {pages} {081201(R)} (\bibinfo {year} {2015})}\BibitemShut
  {NoStop}%
\bibitem [{\citenamefont {Rojas~S{\'a}nchez}\ \emph {et~al.}(2013)\citenamefont
  {Rojas~S{\'a}nchez}, \citenamefont {Vila}, \citenamefont {Desfonds},
  \citenamefont {Gambarelli}, \citenamefont {Attan{\'e}}, \citenamefont
  {De~Teresa}, \citenamefont {Mag{\'e}n},\ and\ \citenamefont
  {Fert}}]{sanchez2013spin}%
  \BibitemOpen
  \bibfield  {author} {\bibinfo {author} {\bibfnamefont {J.~C.}\ \bibnamefont
  {Rojas~S{\'a}nchez}}, \bibinfo {author} {\bibfnamefont {L.}~\bibnamefont
  {Vila}}, \bibinfo {author} {\bibfnamefont {G.}~\bibnamefont {Desfonds}},
  \bibinfo {author} {\bibfnamefont {S.}~\bibnamefont {Gambarelli}}, \bibinfo
  {author} {\bibfnamefont {J.~P.}\ \bibnamefont {Attan{\'e}}}, \bibinfo
  {author} {\bibfnamefont {J.~M.}\ \bibnamefont {De~Teresa}}, \bibinfo {author}
  {\bibfnamefont {C.}~\bibnamefont {Mag{\'e}n}}, \ and\ \bibinfo {author}
  {\bibfnamefont {A.}~\bibnamefont {Fert}},\ }\href@noop {} {\bibfield
  {journal} {\bibinfo  {journal} {Nat. Commun.}\ }\textbf {\bibinfo {volume}
  {4}},\ \bibinfo {pages} {2944} (\bibinfo {year} {2013})}\BibitemShut
  {NoStop}%
\bibitem [{\citenamefont {Lesne}\ \emph {et~al.}(2016)\citenamefont {Lesne},
  \citenamefont {Fu}, \citenamefont {Oyarzun}, \citenamefont
  {Rojas~S{\'a}nchez}, \citenamefont {Vaz}, \citenamefont {Naganuma},
  \citenamefont {Sicoli}, \citenamefont {Attan{\'e}}, \citenamefont {Jamet},
  \citenamefont {Jacquet} \emph {et~al.}}]{lesne2016highly}%
  \BibitemOpen
  \bibfield  {author} {\bibinfo {author} {\bibfnamefont {E.}~\bibnamefont
  {Lesne}}, \bibinfo {author} {\bibfnamefont {Y.}~\bibnamefont {Fu}}, \bibinfo
  {author} {\bibfnamefont {S.}~\bibnamefont {Oyarzun}}, \bibinfo {author}
  {\bibfnamefont {J.~C.}\ \bibnamefont {Rojas~S{\'a}nchez}}, \bibinfo {author}
  {\bibfnamefont {D.~C.}\ \bibnamefont {Vaz}}, \bibinfo {author} {\bibfnamefont
  {H.}~\bibnamefont {Naganuma}}, \bibinfo {author} {\bibfnamefont
  {G.}~\bibnamefont {Sicoli}}, \bibinfo {author} {\bibfnamefont {J.-P.}\
  \bibnamefont {Attan{\'e}}}, \bibinfo {author} {\bibfnamefont
  {M.}~\bibnamefont {Jamet}}, \bibinfo {author} {\bibfnamefont
  {E.}~\bibnamefont {Jacquet}},  \emph {et~al.},\ }\href@noop {} {\bibfield
  {journal} {\bibinfo  {journal} {Nat. Mater.}\ }\textbf {\bibinfo {volume}
  {15}},\ \bibinfo {pages} {1261} (\bibinfo {year} {2016})}\BibitemShut
  {NoStop}%
\bibitem [{\citenamefont {Oyarz{\'u}n}\ \emph {et~al.}(2016)\citenamefont
  {Oyarz{\'u}n}, \citenamefont {Nandy}, \citenamefont {Rortais}, \citenamefont
  {Rojas-S{\'a}nchez}, \citenamefont {Dau}, \citenamefont {No{\"e}l},
  \citenamefont {Laczkowski}, \citenamefont {Pouget}, \citenamefont {Okuno},
  \citenamefont {Vila} \emph {et~al.}}]{oyarzun2016evidence}%
  \BibitemOpen
  \bibfield  {author} {\bibinfo {author} {\bibfnamefont {S.}~\bibnamefont
  {Oyarz{\'u}n}}, \bibinfo {author} {\bibfnamefont {A.~K.}\ \bibnamefont
  {Nandy}}, \bibinfo {author} {\bibfnamefont {F.}~\bibnamefont {Rortais}},
  \bibinfo {author} {\bibfnamefont {J.~C.}\ \bibnamefont {Rojas-S{\'a}nchez}},
  \bibinfo {author} {\bibfnamefont {M.~T.}\ \bibnamefont {Dau}}, \bibinfo
  {author} {\bibfnamefont {P.}~\bibnamefont {No{\"e}l}}, \bibinfo {author}
  {\bibfnamefont {P.}~\bibnamefont {Laczkowski}}, \bibinfo {author}
  {\bibfnamefont {S.}~\bibnamefont {Pouget}}, \bibinfo {author} {\bibfnamefont
  {H.}~\bibnamefont {Okuno}}, \bibinfo {author} {\bibfnamefont
  {L.}~\bibnamefont {Vila}},  \emph {et~al.},\ }\href@noop {} {\bibfield
  {journal} {\bibinfo  {journal} {Nat. Commun.}\ }\textbf {\bibinfo {volume}
  {7}},\ \bibinfo {pages} {13857} (\bibinfo {year} {2016})}\BibitemShut
  {NoStop}%
\bibitem [{\citenamefont {Soumyanarayanan}\ \emph {et~al.}(2016)\citenamefont
  {Soumyanarayanan}, \citenamefont {Reyren}, \citenamefont {Fert},\ and\
  \citenamefont {Panagopoulos}}]{soumyanarayanan2016emergent}%
  \BibitemOpen
  \bibfield  {author} {\bibinfo {author} {\bibfnamefont {A.}~\bibnamefont
  {Soumyanarayanan}}, \bibinfo {author} {\bibfnamefont {N.}~\bibnamefont
  {Reyren}}, \bibinfo {author} {\bibfnamefont {A.}~\bibnamefont {Fert}}, \ and\
  \bibinfo {author} {\bibfnamefont {C.}~\bibnamefont {Panagopoulos}},\
  }\href@noop {} {\bibfield  {journal} {\bibinfo  {journal} {Nature}\ }\textbf
  {\bibinfo {volume} {539}},\ \bibinfo {pages} {509} (\bibinfo {year}
  {2016})}\BibitemShut {NoStop}%
\bibitem [{\citenamefont {Rinaldi}\ \emph {et~al.}(2016)\citenamefont
  {Rinaldi}, \citenamefont {Rojas~S{\'a}nchez}, \citenamefont {Wang},
  \citenamefont {Fu}, \citenamefont {Oyarzun}, \citenamefont {Vila},
  \citenamefont {Bertoli}, \citenamefont {Asa}, \citenamefont {Baldrati},
  \citenamefont {Cantoni} \emph {et~al.}}]{rinaldi2016evidence}%
  \BibitemOpen
  \bibfield  {author} {\bibinfo {author} {\bibfnamefont {C.}~\bibnamefont
  {Rinaldi}}, \bibinfo {author} {\bibfnamefont {J.~C.}\ \bibnamefont
  {Rojas~S{\'a}nchez}}, \bibinfo {author} {\bibfnamefont {R.~N.}\ \bibnamefont
  {Wang}}, \bibinfo {author} {\bibfnamefont {Y.}~\bibnamefont {Fu}}, \bibinfo
  {author} {\bibfnamefont {S.}~\bibnamefont {Oyarzun}}, \bibinfo {author}
  {\bibfnamefont {L.}~\bibnamefont {Vila}}, \bibinfo {author} {\bibfnamefont
  {S.}~\bibnamefont {Bertoli}}, \bibinfo {author} {\bibfnamefont
  {M.}~\bibnamefont {Asa}}, \bibinfo {author} {\bibfnamefont {L.}~\bibnamefont
  {Baldrati}}, \bibinfo {author} {\bibfnamefont {M.}~\bibnamefont {Cantoni}},
  \emph {et~al.},\ }\href@noop {} {\bibfield  {journal} {\bibinfo  {journal}
  {APL Mater.}\ }\textbf {\bibinfo {volume} {4}},\ \bibinfo {pages} {032501}
  (\bibinfo {year} {2016})}\BibitemShut {NoStop}%
\bibitem [{\citenamefont {Di~Sante}\ \emph {et~al.}(2013)\citenamefont
  {Di~Sante}, \citenamefont {Barone}, \citenamefont {Bertacco},\ and\
  \citenamefont {Picozzi}}]{di2013electric}%
  \BibitemOpen
  \bibfield  {author} {\bibinfo {author} {\bibfnamefont {D.}~\bibnamefont
  {Di~Sante}}, \bibinfo {author} {\bibfnamefont {P.}~\bibnamefont {Barone}},
  \bibinfo {author} {\bibfnamefont {R.}~\bibnamefont {Bertacco}}, \ and\
  \bibinfo {author} {\bibfnamefont {S.}~\bibnamefont {Picozzi}},\ }\href@noop
  {} {\bibfield  {journal} {\bibinfo  {journal} {Adv. Mater.}\ }\textbf
  {\bibinfo {volume} {25}},\ \bibinfo {pages} {509} (\bibinfo {year}
  {2013})}\BibitemShut {NoStop}%
\bibitem [{\citenamefont {Picozzi}(2014)}]{picozzi2014ferroelectric}%
  \BibitemOpen
  \bibfield  {author} {\bibinfo {author} {\bibfnamefont {S.}~\bibnamefont
  {Picozzi}},\ }\href@noop {} {\bibfield  {journal} {\bibinfo  {journal}
  {Front. Phys.}\ }\textbf {\bibinfo {volume} {2}},\ \bibinfo {pages} {10}
  (\bibinfo {year} {2014})}\BibitemShut {NoStop}%
\bibitem [{\citenamefont {Liebmann}\ \emph {et~al.}(2016)\citenamefont
  {Liebmann}, \citenamefont {Rinaldi}, \citenamefont {Di~Sante}, \citenamefont
  {Kellner}, \citenamefont {Pauly}, \citenamefont {Wang}, \citenamefont
  {Boschker}, \citenamefont {Giussani}, \citenamefont {Bertoli}, \citenamefont
  {Cantoni}, \citenamefont {Baldrati}, \citenamefont {Asa}, \citenamefont
  {Vobornik}, \citenamefont {Panaccione}, \citenamefont {Marchenko},
  \citenamefont {S{\'a}nchez-Bariga}, \citenamefont {Rader}, \citenamefont
  {Calarco}, \citenamefont {Picozzi}, \citenamefont {Bertacco},\ and\
  \citenamefont {Morgenstern}}]{liebmann2016giant}%
  \BibitemOpen
  \bibfield  {author} {\bibinfo {author} {\bibfnamefont {M.}~\bibnamefont
  {Liebmann}}, \bibinfo {author} {\bibfnamefont {C.}~\bibnamefont {Rinaldi}},
  \bibinfo {author} {\bibfnamefont {D.}~\bibnamefont {Di~Sante}}, \bibinfo
  {author} {\bibfnamefont {J.}~\bibnamefont {Kellner}}, \bibinfo {author}
  {\bibfnamefont {C.}~\bibnamefont {Pauly}}, \bibinfo {author} {\bibfnamefont
  {R.~N.}\ \bibnamefont {Wang}}, \bibinfo {author} {\bibfnamefont {J.~E.}\
  \bibnamefont {Boschker}}, \bibinfo {author} {\bibfnamefont {A.}~\bibnamefont
  {Giussani}}, \bibinfo {author} {\bibfnamefont {S.}~\bibnamefont {Bertoli}},
  \bibinfo {author} {\bibfnamefont {M.}~\bibnamefont {Cantoni}}, \bibinfo
  {author} {\bibfnamefont {L.}~\bibnamefont {Baldrati}}, \bibinfo {author}
  {\bibfnamefont {M.}~\bibnamefont {Asa}}, \bibinfo {author} {\bibfnamefont
  {I.}~\bibnamefont {Vobornik}}, \bibinfo {author} {\bibfnamefont
  {G.}~\bibnamefont {Panaccione}}, \bibinfo {author} {\bibfnamefont
  {D.}~\bibnamefont {Marchenko}}, \bibinfo {author} {\bibfnamefont
  {J.}~\bibnamefont {S{\'a}nchez-Bariga}}, \bibinfo {author} {\bibfnamefont
  {O.}~\bibnamefont {Rader}}, \bibinfo {author} {\bibfnamefont
  {R.}~\bibnamefont {Calarco}}, \bibinfo {author} {\bibfnamefont
  {S.}~\bibnamefont {Picozzi}}, \bibinfo {author} {\bibfnamefont
  {R.}~\bibnamefont {Bertacco}}, \ and\ \bibinfo {author} {\bibfnamefont
  {M.}~\bibnamefont {Morgenstern}},\ }\href@noop {} {\bibfield  {journal}
  {\bibinfo  {journal} {Adv. Mater.}\ }\textbf {\bibinfo {volume} {28}},\
  \bibinfo {pages} {560} (\bibinfo {year} {2016})}\BibitemShut {NoStop}%
\bibitem [{\citenamefont {Krempask{\'y}}\ \emph
  {et~al.}(2016{\natexlab{a}})\citenamefont {Krempask{\'y}}, \citenamefont
  {Muff}, \citenamefont {Bisti}, \citenamefont {Fanciulli}, \citenamefont
  {Volfov{\'a}}, \citenamefont {Weber}, \citenamefont {Pilet}, \citenamefont
  {Warnicke}, \citenamefont {Ebert}, \citenamefont {Braun}, \citenamefont
  {Bertran}, \citenamefont {Volobuev}, \citenamefont {Min{\'a}r}, \citenamefont
  {Springholz}, \citenamefont {Dil},\ and\ \citenamefont
  {Strocov}}]{krempasky2016entanglement}%
  \BibitemOpen
  \bibfield  {author} {\bibinfo {author} {\bibfnamefont {J.}~\bibnamefont
  {Krempask{\'y}}}, \bibinfo {author} {\bibfnamefont {S.}~\bibnamefont {Muff}},
  \bibinfo {author} {\bibfnamefont {F.}~\bibnamefont {Bisti}}, \bibinfo
  {author} {\bibfnamefont {M.}~\bibnamefont {Fanciulli}}, \bibinfo {author}
  {\bibfnamefont {H.}~\bibnamefont {Volfov{\'a}}}, \bibinfo {author}
  {\bibfnamefont {A.~P.}\ \bibnamefont {Weber}}, \bibinfo {author}
  {\bibfnamefont {N.}~\bibnamefont {Pilet}}, \bibinfo {author} {\bibfnamefont
  {P.}~\bibnamefont {Warnicke}}, \bibinfo {author} {\bibfnamefont
  {H.}~\bibnamefont {Ebert}}, \bibinfo {author} {\bibfnamefont
  {J.}~\bibnamefont {Braun}}, \bibinfo {author} {\bibfnamefont
  {F.}~\bibnamefont {Bertran}}, \bibinfo {author} {\bibfnamefont {V.~V.}\
  \bibnamefont {Volobuev}}, \bibinfo {author} {\bibfnamefont {J.}~\bibnamefont
  {Min{\'a}r}}, \bibinfo {author} {\bibfnamefont {G.}~\bibnamefont
  {Springholz}}, \bibinfo {author} {\bibfnamefont {J.~H.}\ \bibnamefont {Dil}},
  \ and\ \bibinfo {author} {\bibfnamefont {V.~N.}\ \bibnamefont {Strocov}},\
  }\href@noop {} {\bibfield  {journal} {\bibinfo  {journal} {Nat. Commun.}\
  }\textbf {\bibinfo {volume} {7}},\ \bibinfo {pages} {13071} (\bibinfo {year}
  {2016}{\natexlab{a}})}\BibitemShut {NoStop}%
\bibitem [{\citenamefont {Krempask{\'y}}\ \emph
  {et~al.}(2016{\natexlab{b}})\citenamefont {Krempask{\'y}}, \citenamefont
  {Volfov{\'a}}, \citenamefont {Muff}, \citenamefont {Pilet}, \citenamefont
  {Landolt}, \citenamefont {Radovi{\'c}}, \citenamefont {Shi}, \citenamefont
  {Kriegner}, \citenamefont {Hol{\`y}}, \citenamefont {Braun}, \citenamefont
  {Ebert}, \citenamefont {Bisti}, \citenamefont {Rogalev}, \citenamefont
  {Strocov}, \citenamefont {Springholz}, \citenamefont {Min{\'a}r},\ and\
  \citenamefont {Dil}}]{krempasky2016disentangling}%
  \BibitemOpen
  \bibfield  {author} {\bibinfo {author} {\bibfnamefont {J.}~\bibnamefont
  {Krempask{\'y}}}, \bibinfo {author} {\bibfnamefont {H.}~\bibnamefont
  {Volfov{\'a}}}, \bibinfo {author} {\bibfnamefont {S.}~\bibnamefont {Muff}},
  \bibinfo {author} {\bibfnamefont {N.}~\bibnamefont {Pilet}}, \bibinfo
  {author} {\bibfnamefont {G.}~\bibnamefont {Landolt}}, \bibinfo {author}
  {\bibfnamefont {M.}~\bibnamefont {Radovi{\'c}}}, \bibinfo {author}
  {\bibfnamefont {M.}~\bibnamefont {Shi}}, \bibinfo {author} {\bibfnamefont
  {D.}~\bibnamefont {Kriegner}}, \bibinfo {author} {\bibfnamefont
  {V.}~\bibnamefont {Hol{\`y}}}, \bibinfo {author} {\bibfnamefont
  {J.}~\bibnamefont {Braun}}, \bibinfo {author} {\bibfnamefont
  {H.}~\bibnamefont {Ebert}}, \bibinfo {author} {\bibfnamefont
  {F.}~\bibnamefont {Bisti}}, \bibinfo {author} {\bibfnamefont {V.~A.}\
  \bibnamefont {Rogalev}}, \bibinfo {author} {\bibfnamefont {V.~N.}\
  \bibnamefont {Strocov}}, \bibinfo {author} {\bibfnamefont {G.}~\bibnamefont
  {Springholz}}, \bibinfo {author} {\bibfnamefont {J.}~\bibnamefont
  {Min{\'a}r}}, \ and\ \bibinfo {author} {\bibfnamefont {J.~H.}\ \bibnamefont
  {Dil}},\ }\href@noop {} {\bibfield  {journal} {\bibinfo  {journal} {Phys.
  Rev. B}\ }\textbf {\bibinfo {volume} {94}},\ \bibinfo {pages} {205111}
  (\bibinfo {year} {2016}{\natexlab{b}})}\BibitemShut {NoStop}%
\bibitem [{\citenamefont {Rinaldi}\ \emph {et~al.}(2018)\citenamefont
  {Rinaldi}, \citenamefont {Varotto}, \citenamefont {Asa}, \citenamefont
  {S{\l}awin{\'n}ska}, \citenamefont {Fujii}, \citenamefont {Vinai},
  \citenamefont {Cecchi}, \citenamefont {Di~Sante}, \citenamefont {Calarco},
  \citenamefont {Vobornik}, \citenamefont {Panaccione}, \citenamefont
  {Picozzi},\ and\ \citenamefont {Bertacco}}]{rinaldi2018ferroelectric}%
  \BibitemOpen
  \bibfield  {author} {\bibinfo {author} {\bibfnamefont {C.}~\bibnamefont
  {Rinaldi}}, \bibinfo {author} {\bibfnamefont {S.}~\bibnamefont {Varotto}},
  \bibinfo {author} {\bibfnamefont {M.}~\bibnamefont {Asa}}, \bibinfo {author}
  {\bibfnamefont {J.}~\bibnamefont {S{\l}awin{\'n}ska}}, \bibinfo {author}
  {\bibfnamefont {J.}~\bibnamefont {Fujii}}, \bibinfo {author} {\bibfnamefont
  {G.}~\bibnamefont {Vinai}}, \bibinfo {author} {\bibfnamefont
  {S.}~\bibnamefont {Cecchi}}, \bibinfo {author} {\bibfnamefont
  {D.}~\bibnamefont {Di~Sante}}, \bibinfo {author} {\bibfnamefont
  {R.}~\bibnamefont {Calarco}}, \bibinfo {author} {\bibfnamefont
  {I.}~\bibnamefont {Vobornik}}, \bibinfo {author} {\bibfnamefont
  {G.}~\bibnamefont {Panaccione}}, \bibinfo {author} {\bibfnamefont
  {S.}~\bibnamefont {Picozzi}}, \ and\ \bibinfo {author} {\bibfnamefont
  {R.}~\bibnamefont {Bertacco}},\ }\href@noop {} {\bibfield  {journal}
  {\bibinfo  {journal} {Nano Lett.}\ }\textbf {\bibinfo {volume} {18}},\
  \bibinfo {pages} {2751} (\bibinfo {year} {2018})}\BibitemShut {NoStop}%
\bibitem [{\citenamefont {Krempask{\'y}}\ \emph {et~al.}(2018)\citenamefont
  {Krempask{\'y}}, \citenamefont {Muff}, \citenamefont {Minar}, \citenamefont
  {Pilet}, \citenamefont {Fanciulli}, \citenamefont {Weber}, \citenamefont
  {Guedes}, \citenamefont {Caputo}, \citenamefont {M{\"u}ller}, \citenamefont
  {Volobuev}, \citenamefont {Gmitra}, \citenamefont {Vaz}, \citenamefont
  {Scagnoli}, \citenamefont {Springholz},\ and\ \citenamefont
  {Dil}}]{krempasky2018operando}%
  \BibitemOpen
  \bibfield  {author} {\bibinfo {author} {\bibfnamefont {J.}~\bibnamefont
  {Krempask{\'y}}}, \bibinfo {author} {\bibfnamefont {S.}~\bibnamefont {Muff}},
  \bibinfo {author} {\bibfnamefont {J.}~\bibnamefont {Minar}}, \bibinfo
  {author} {\bibfnamefont {N.}~\bibnamefont {Pilet}}, \bibinfo {author}
  {\bibfnamefont {M.}~\bibnamefont {Fanciulli}}, \bibinfo {author}
  {\bibfnamefont {A.~P.}\ \bibnamefont {Weber}}, \bibinfo {author}
  {\bibfnamefont {E.~B.}\ \bibnamefont {Guedes}}, \bibinfo {author}
  {\bibfnamefont {M.}~\bibnamefont {Caputo}}, \bibinfo {author} {\bibfnamefont
  {E.}~\bibnamefont {M{\"u}ller}}, \bibinfo {author} {\bibfnamefont {V.~V.}\
  \bibnamefont {Volobuev}}, \bibinfo {author} {\bibfnamefont {M.}~\bibnamefont
  {Gmitra}}, \bibinfo {author} {\bibfnamefont {C.~A.~F.}\ \bibnamefont {Vaz}},
  \bibinfo {author} {\bibfnamefont {V.}~\bibnamefont {Scagnoli}}, \bibinfo
  {author} {\bibfnamefont {G.}~\bibnamefont {Springholz}}, \ and\ \bibinfo
  {author} {\bibfnamefont {J.~H.}\ \bibnamefont {Dil}},\ }\href@noop {}
  {\bibfield  {journal} {\bibinfo  {journal} {Phys. Rev. X}\ }\textbf {\bibinfo
  {volume} {8}},\ \bibinfo {pages} {021067} (\bibinfo {year}
  {2018})}\BibitemShut {NoStop}%
\bibitem [{\citenamefont {Kremer}\ \emph {et~al.}(2019)\citenamefont {Kremer},
  \citenamefont {Zhu}, \citenamefont {Pierron}, \citenamefont {Fourn{\'e}e},
  \citenamefont {Ledieu}, \citenamefont {Andrieu}, \citenamefont {Kierren},
  \citenamefont {Moreau}, \citenamefont {Malterre}, \citenamefont {He},
  \citenamefont {Xue}, \citenamefont {Fagot-Revurat},\ and\ \citenamefont
  {Lu}}]{kremer2019recovery}%
  \BibitemOpen
  \bibfield  {author} {\bibinfo {author} {\bibfnamefont {G.}~\bibnamefont
  {Kremer}}, \bibinfo {author} {\bibfnamefont {K.}~\bibnamefont {Zhu}},
  \bibinfo {author} {\bibfnamefont {T.}~\bibnamefont {Pierron}}, \bibinfo
  {author} {\bibfnamefont {V.}~\bibnamefont {Fourn{\'e}e}}, \bibinfo {author}
  {\bibfnamefont {J.}~\bibnamefont {Ledieu}}, \bibinfo {author} {\bibfnamefont
  {S.}~\bibnamefont {Andrieu}}, \bibinfo {author} {\bibfnamefont
  {B.}~\bibnamefont {Kierren}}, \bibinfo {author} {\bibfnamefont
  {L.}~\bibnamefont {Moreau}}, \bibinfo {author} {\bibfnamefont
  {D.}~\bibnamefont {Malterre}}, \bibinfo {author} {\bibfnamefont
  {K.}~\bibnamefont {He}}, \bibinfo {author} {\bibfnamefont {Q.~K.}\
  \bibnamefont {Xue}}, \bibinfo {author} {\bibfnamefont {Y.}~\bibnamefont
  {Fagot-Revurat}}, \ and\ \bibinfo {author} {\bibfnamefont {Y.}~\bibnamefont
  {Lu}},\ }\href@noop {} {\bibfield  {journal} {\bibinfo  {journal} {J. Phys.
  D: Appl. Phys.}\ }\textbf {\bibinfo {volume} {52}},\ \bibinfo {pages}
  {494002} (\bibinfo {year} {2019})}\BibitemShut {NoStop}%
\bibitem [{\citenamefont {Ebert}\ \emph {et~al.}(2011)\citenamefont {Ebert},
  \citenamefont {K{\"o}dderitzsch},\ and\ \citenamefont
  {Min{\'ar}}}]{ebert2011calculating}%
  \BibitemOpen
  \bibfield  {author} {\bibinfo {author} {\bibfnamefont {H.}~\bibnamefont
  {Ebert}}, \bibinfo {author} {\bibfnamefont {D.}~\bibnamefont
  {K{\"o}dderitzsch}}, \ and\ \bibinfo {author} {\bibfnamefont
  {J.}~\bibnamefont {Min{\'ar}}},\ }\href@noop {} {\bibfield  {journal}
  {\bibinfo  {journal} {Rep. Prog. Phys.}\ }\textbf {\bibinfo {volume} {74}},\
  \bibinfo {pages} {096501} (\bibinfo {year} {2011})}\BibitemShut {NoStop}%
\bibitem [{\citenamefont {Bauer~Pereira}\ \emph {et~al.}(2013)\citenamefont
  {Bauer~Pereira}, \citenamefont {Sergueev}, \citenamefont {Gorsse},
  \citenamefont {Dadda}, \citenamefont {M{\"u}ller},\ and\ \citenamefont
  {Hermann}}]{bauer2013lattice}%
  \BibitemOpen
  \bibfield  {author} {\bibinfo {author} {\bibfnamefont {P.}~\bibnamefont
  {Bauer~Pereira}}, \bibinfo {author} {\bibfnamefont {I.}~\bibnamefont
  {Sergueev}}, \bibinfo {author} {\bibfnamefont {S.}~\bibnamefont {Gorsse}},
  \bibinfo {author} {\bibfnamefont {J.}~\bibnamefont {Dadda}}, \bibinfo
  {author} {\bibfnamefont {E.}~\bibnamefont {M{\"u}ller}}, \ and\ \bibinfo
  {author} {\bibfnamefont {R.~P.}\ \bibnamefont {Hermann}},\ }\href@noop {}
  {\bibfield  {journal} {\bibinfo  {journal} {Phys. Status Solidi B}\ }\textbf
  {\bibinfo {volume} {250}},\ \bibinfo {pages} {1300} (\bibinfo {year}
  {2013})}\BibitemShut {NoStop}%
\bibitem [{\citenamefont {Deringer}\ \emph {et~al.}(2012)\citenamefont
  {Deringer}, \citenamefont {Lumeij},\ and\ \citenamefont
  {Dronskowski}}]{deringer2012ab}%
  \BibitemOpen
  \bibfield  {author} {\bibinfo {author} {\bibfnamefont {V.~L.}\ \bibnamefont
  {Deringer}}, \bibinfo {author} {\bibfnamefont {M.}~\bibnamefont {Lumeij}}, \
  and\ \bibinfo {author} {\bibfnamefont {R.}~\bibnamefont {Dronskowski}},\
  }\href@noop {} {\bibfield  {journal} {\bibinfo  {journal} {J. Phys. Chem. C}\
  }\textbf {\bibinfo {volume} {116}},\ \bibinfo {pages} {15801} (\bibinfo
  {year} {2012})}\BibitemShut {NoStop}%
\bibitem [{\citenamefont {Krempask{\'y}}\ \emph {et~al.}(2020)\citenamefont
  {Krempask{\'y}}, \citenamefont {Fanciulli}, \citenamefont {Nicola\"{\i}},
  \citenamefont {Min\'ar}, \citenamefont {Volfov\'a}, \citenamefont {Caha},
  \citenamefont {Volobuev}, \citenamefont {S\'anchez-Barriga}, \citenamefont
  {Gmitra}, \citenamefont {Yaji}, \citenamefont {Kuroda}, \citenamefont {Shin},
  \citenamefont {Komori}, \citenamefont {Springholz},\ and\ \citenamefont
  {Dil}}]{PhysResearchKrempasky2020}%
  \BibitemOpen
  \bibfield  {author} {\bibinfo {author} {\bibfnamefont {J.}~\bibnamefont
  {Krempask{\'y}}}, \bibinfo {author} {\bibfnamefont {M.}~\bibnamefont
  {Fanciulli}}, \bibinfo {author} {\bibfnamefont {L.}~\bibnamefont
  {Nicola\"{\i}}}, \bibinfo {author} {\bibfnamefont {J.}~\bibnamefont
  {Min\'ar}}, \bibinfo {author} {\bibfnamefont {H.}~\bibnamefont {Volfov\'a}},
  \bibinfo {author} {\bibfnamefont {O.}~\bibnamefont {Caha}}, \bibinfo {author}
  {\bibfnamefont {V.~V.}\ \bibnamefont {Volobuev}}, \bibinfo {author}
  {\bibfnamefont {J.}~\bibnamefont {S\'anchez-Barriga}}, \bibinfo {author}
  {\bibfnamefont {M.}~\bibnamefont {Gmitra}}, \bibinfo {author} {\bibfnamefont
  {K.}~\bibnamefont {Yaji}}, \bibinfo {author} {\bibfnamefont {K.}~\bibnamefont
  {Kuroda}}, \bibinfo {author} {\bibfnamefont {S.}~\bibnamefont {Shin}},
  \bibinfo {author} {\bibfnamefont {F.}~\bibnamefont {Komori}}, \bibinfo
  {author} {\bibfnamefont {G.}~\bibnamefont {Springholz}}, \ and\ \bibinfo
  {author} {\bibfnamefont {J.~H.}\ \bibnamefont {Dil}},\ }\href@noop {}
  {\bibfield  {journal} {\bibinfo  {journal} {Phys. Rev. Research}\ }\textbf
  {\bibinfo {volume} {2}},\ \bibinfo {pages} {013107} (\bibinfo {year}
  {2020})}\BibitemShut {NoStop}%
\bibitem [{\citenamefont {Hossain}\ \emph {et~al.}(2008)\citenamefont
  {Hossain}, \citenamefont {Mottershead}, \citenamefont {Fournier},
  \citenamefont {Bostwick}, \citenamefont {McChesney}, \citenamefont
  {Rotenberg}, \citenamefont {Liang}, \citenamefont {Hardy}, \citenamefont
  {Sawatzky}, \citenamefont {Elfimov} \emph {et~al.}}]{hossain2008situ}%
  \BibitemOpen
  \bibfield  {author} {\bibinfo {author} {\bibfnamefont {M.~A.}\ \bibnamefont
  {Hossain}}, \bibinfo {author} {\bibfnamefont {J.~D.~F.}\ \bibnamefont
  {Mottershead}}, \bibinfo {author} {\bibfnamefont {D.}~\bibnamefont
  {Fournier}}, \bibinfo {author} {\bibfnamefont {A.}~\bibnamefont {Bostwick}},
  \bibinfo {author} {\bibfnamefont {J.~L.}\ \bibnamefont {McChesney}}, \bibinfo
  {author} {\bibfnamefont {E.}~\bibnamefont {Rotenberg}}, \bibinfo {author}
  {\bibfnamefont {R.}~\bibnamefont {Liang}}, \bibinfo {author} {\bibfnamefont
  {W.~N.}\ \bibnamefont {Hardy}}, \bibinfo {author} {\bibfnamefont {G.~A.}\
  \bibnamefont {Sawatzky}}, \bibinfo {author} {\bibfnamefont {I.~S.}\
  \bibnamefont {Elfimov}},  \emph {et~al.},\ }\href@noop {} {\bibfield
  {journal} {\bibinfo  {journal} {Nat. Phys.}\ }\textbf {\bibinfo {volume}
  {4}},\ \bibinfo {pages} {527} (\bibinfo {year} {2008})}\BibitemShut {NoStop}%
\bibitem [{\citenamefont {Boyle}\ \emph {et~al.}(2019)\citenamefont {Boyle},
  \citenamefont {Rossi}, \citenamefont {Walker}, \citenamefont {Carlson},
  \citenamefont {Miller}, \citenamefont {Zhao}, \citenamefont {Klavins},
  \citenamefont {Jozwiak}, \citenamefont {Bostwick}, \citenamefont {Rotenberg}
  \emph {et~al.}}]{boyle2019topological}%
  \BibitemOpen
  \bibfield  {author} {\bibinfo {author} {\bibfnamefont {T.~J.}\ \bibnamefont
  {Boyle}}, \bibinfo {author} {\bibfnamefont {A.}~\bibnamefont {Rossi}},
  \bibinfo {author} {\bibfnamefont {M.}~\bibnamefont {Walker}}, \bibinfo
  {author} {\bibfnamefont {P.}~\bibnamefont {Carlson}}, \bibinfo {author}
  {\bibfnamefont {M.~K.}\ \bibnamefont {Miller}}, \bibinfo {author}
  {\bibfnamefont {J.}~\bibnamefont {Zhao}}, \bibinfo {author} {\bibfnamefont
  {P.}~\bibnamefont {Klavins}}, \bibinfo {author} {\bibfnamefont
  {C.}~\bibnamefont {Jozwiak}}, \bibinfo {author} {\bibfnamefont
  {A.}~\bibnamefont {Bostwick}}, \bibinfo {author} {\bibfnamefont
  {E.}~\bibnamefont {Rotenberg}},  \emph {et~al.},\ }\href@noop {} {\bibfield
  {journal} {\bibinfo  {journal} {Phys. Rev. B}\ }\textbf {\bibinfo {volume}
  {100}},\ \bibinfo {pages} {081105(R)} (\bibinfo {year} {2019})}\BibitemShut
  {NoStop}%
\end{thebibliography}
\end{document}